\documentclass[fleqn,usenatbib]{mnras}
\usepackage{newtxtext,newtxmath}
\usepackage[T1]{fontenc}
\DeclareRobustCommand{\VAN}[3]{#2}
\let\VANthebibliography\thebibliography
\def\thebibliography{\DeclareRobustCommand{\VAN}[3]{##3}\VANthebibliography}
\usepackage{graphicx}	
\usepackage{amsmath}
\usepackage{ulem}

\title{On the Binary Nature of the Progenitor of SN2015ap: Insights from Its Light Curve and Spectral Evolution}

\author[Ragosta F. et al]{
Fabio Ragosta$^{1,2}$\thanks{E-mail: fabio.ragosta@unina.it},
Giulia Illiano$^{3}$,
Andrea Simongini$^{4,5}$,
\'Osmar Rodr\'iguez$^{6,7}$,
Matteo Imbrogno$^{4}$,\newauthor
Silvia Piranomonte$^{4}$,
Andrea Melandri$^{4}$
\\
$^{1}$Dipartimento di Fisica “Ettore Pancini”, Università di Napoli Federico II, Via Cinthia 9, 80126 Naples, Italy \\
$^{2}$INAF - Osservatorio Astronomico di Capodimonte, Via Moiariello 16, I-80131 Naples, Italy\\
$^{3}$ INAF - Osservatorio Astronomico di Brera, Via Bianchi 46, I-23807, Merate (LC), Italy\\
$^{4}$INAF - Osservatorio Astronomico di Roma, Via di Frascati 33, I-00078 Monteporzio Catone, Italy\\
$^{5}$Università Tor Vergata, Dipartimento di Fisica, Via della Ricerca Scientifica 1, I-00133 Rome, Italy\\ 
$^{6}$ Pontiﬁcia Universidad Cat\'olica de Chile, Vicu\~na Mackenna 4860, Macul, Santiago, Chile\\
$^{7}$ Instituto Milenio de Astrof\'isica (MAS), Nuncio Monse\~nor S\'otero Sanz 100, Of. 104, Santiago, Chile
}

\date{Accepted XXX. Received YYY; in original form ZZZ}

\pubyear{\the\year{}}

\begin{document}
\label{firstpage}
\pagerange{\pageref{firstpage}--\pageref{lastpage}}
\maketitle

\begin{abstract}
Stripped-envelope supernovae (SESNe) display a wide range of photometric and spectroscopic behaviours, often reflecting complex progenitor evolution. SN~2015ap is a type Ib event located in the nearby galaxy IC~1776, previously modelled as powered by radioactive decay and possibly a magnetar engine. In this work, we revisit its multi-band photometry and spectroscopy, {gathering all publicly available observational data for this source}, to investigate the nature of its progenitor and power source.
{We use an innovative time analysis method based on Gaussian Process, leveraging its ability to model both noise and periodic components in unevenly sampled data without requiring regular sampling.}
We detect significant periodic modulations in the post-peak light curve, with a characteristic timescale of $\sim$8.4 days. These modulations are also seen in the $H_{\alpha}$ line velocity, suggesting a structured circumstellar medium (CSM) shaped by binary interaction. We model the light curve with semi-analytical prescriptions (MOSFiT), including CSM and central engine components, and derive an ejecta mass of $\sim$2.2--2.4~$M_\odot$, explosion energy of $\sim$3.4$\times$10$^{51}$~erg, and a $^{56}$Ni mass of $\sim$0.11~$M_\odot$. The colour evolution indicates an additional energy injection, consistent with either prolonged breakout or delayed central powering. While the data are compatible with a weak magnetar contribution, the overall evidence favours a binary progenitor system, with non-conservative mass transfer shaping the observed CSM. SN~2015ap thus adds to the growing sample of SESNe where binarity plays a central role in driving both the explosion and its observables.
\end{abstract}

\begin{keywords}
(stars:) binaries (including multiple): close < Stars, (stars:) supernovae:
general < Stars, transients: supernovae < Transients, (stars:)
circumstellar matter < Stars
\end{keywords}

\section{Introduction}

Supernovae (SNe) are among the most energetic transient phenomena in the Universe. They arise from two primary mechanisms: the gravitational collapse of massive stars with initial mass $M \gtrsim 8\,M_{\odot}$ (core-collapse SNe, CCSNe), and thermonuclear explosions of carbon-oxygen white dwarfs in close binary systems (type Ia SNe). In both cases, the explosion releases an energy of order $10^{51}$ erg, although only $\sim1\%$ of this energy is emitted as electromagnetic radiation, which we observe as the SN light curve and spectra \citep{filippenko1997optical, 2005Filippenko}.

SNe are classified according to their early-time spectral features. The primary distinction is the presence or absence of hydrogen lines: type II SNe exhibit prominent hydrogen features, while type I SNe do not. Further subdivisions are made based on other spectral characteristics: while type Ia SNe show a strong Si II absorption feature, type Ib display helium lines, and type Ic lack both helium and silicon features. While type Ia SNe are relatively homogeneous and are used as standard candles in cosmology, CCSNe (types II, IIb, Ib, and Ic) display considerable diversity in both photometric and spectroscopic properties \citep{branch1990supernova, filippenko1997optical, turatto2003classification, galYam2016observational}.

{The lack of hydrogen or helium is often related to mass-loss mechanisms that strip the progenitor star of most or all of its outer envelope before explosion. Because of this, supernovae of types IIb, Ib, and Ic are collectively referred to as stripped-envelope supernovae (SESNe)}

This diversity among CCSNe is driven by a range of physical parameters, including the progenitor’s mass, metallicity, rotation, and especially its mass-loss history. These factors influence the environment into which the SN explodes, thereby shaping the observed light curve and spectral evolution\citep{smartt2009progenitors, woosley2002evolution, yoon2010massive, smith2014mass}.

During the early photospheric phase, SN light curves are typically powered by shock-deposited energy and, later, by radioactive decay, primarily from $^{56}$Ni to $^{56}$Co and then to $^{56}$Fe. However, in many cases, deviations from the expected decay-driven evolution suggest the presence of additional energy sources. One of the most important among these is the interaction between the SN ejecta and circumstellar material (CSM), wherein the collision produces forward and reverse shocks that reprocess kinetic energy into electromagnetic emission. This mechanism can significantly enhance or modify the SN light curve \citep[e.g.,][]{1982Chevalier, 1982Arnett, 1994Chevalier, 2011Moriya, Chatzopoulos_2013, Guillochon_2018, 2024Srinivasaragavan}.

Some SESN, especially of type Ib, show evidence for energy injection that cannot be explained solely by radioactive decay or CSM interaction. These events have been interpreted as being powered by central engines, such as newly born magnetars \citep{2020ApJ...890...51E,2021ApJ...918...89A, 2024Natur.628..733R}, which can substantially modify the luminosity evolution \citep[e.g.,][]{1969Ostriker, arnett1989late, maeda2007unique, Kasen_2010, 2010Woosley, Chatzopoulos_2013, Nicholl}. 

Over the past decade, a growing number of CCSNe have exhibited rebrightenings or undulations (“bumps”) in their light curves during both early and late phases. These features are commonly interpreted as signatures of CSM interaction, but their properties are often inconsistent with the smooth, steady wind mass-loss expected from Wolf–Rayet stars with $M \gtrsim 30\,M_\odot$. For example, the well-known case of SN 1987A reveals a complex circumstellar morphology and episodic mass-loss history that cannot be reproduced by standard wind models \citep{moriya2014light, smith2017observational, arnett1989sn1987a, mccray2016supernova}.

A particularly informative class of features is the early-time bumps observed in some SESNe. These luminousity excesses, occurring within days to weeks of the explosion, are interpreted as evidence of ejecta-CSM interaction shortly after shock breakout. Their photometric and temporal properties offer insights into the density, distribution, and timing of mass loss from the progenitor system. Recent works suggest that the observed CSM in many SESNe may originate not from steady winds or intrinsic outbursts, but rather from binary interaction \citep[see, e.g.,][]{Kuncarayakti}. Mass transfer episodes, tidal effects, and common-envelope evolution in binary systems can produce complex, often asymmetric CSM geometries and significantly enhanced mass-loss rates as proposed by \citet{Bersten_2014} for the case of iPTF13bvn. These processes provide a natural explanation for the presence of dense CSM near the progenitor at the time of explosion. In this context, early-time bumps in SN light curves become powerful diagnostics of binary-induced CSM shaping and offer a unique observational probe of the final stages of binary stellar evolution.

Binary interaction is also crucial in explaining the observed frequency of hydrogen-poor CCSNe (types IIb, Ib, Ic), which exceeds predictions from single-star evolution alone \citep[e.g.,][]{1992Podsiadlowski, 2008Eldridge, yoon2010type, claeys2011binary, 2017Zapartas, 2020Sravan, 2022Shivvers}. Binary evolution scenarios naturally account for the low ejecta masses ($\sim2\,M_\odot$) frequently inferred from light-curve modelling \citep[e.g.,][]{Lyman2016, 2018Taddia}, and are supported by direct observations of SN progenitor environments. For instance, the triple-ring nebula surrounding SN 1987A is now widely interpreted as a product of binary merger or interaction events \citep[e.g.,][]{2015Eldridge, 2017Eldridge, Utrobin_2021}.

Previous studies on SN~2015ap -- a bright type Ib SN discovered in IC~1776 -- have provided detailed photometric and spectroscopic characterisations that already point toward a moderately massive progenitor in a binary system \citep[e.g. ][]{Prentice_2018, 2020Gangopadhyay,Aryan2021, 2022JApA...43...87A}. \citet{2020Gangopadhyay} modelled the bolometric light curve using a hybrid $^{56}$Ni~+~magnetar scenario, finding a nickel mass ($^{56}Ni$) of $\sim$0.01~M$_\odot$, ejecta mass of $\sim$3.75~M$_\odot$, and magnetar spin period of 25.8~ms with a magnetic field strength of $2.8 \times 10^{15}$~G. The early spectra display strong He~\textsc{i} and Fe~\textsc{ii} lines at high velocities ($\sim$15,500~km~s$^{-1}$), with the helium features fading post-maximum as the photosphere cools. In the nebular phase, the [O~\textsc{i}]~$\lambda\lambda$6300,6364 lines show a marked blue-shifted asymmetry, suggesting an aspherical explosion geometry.

The nebular line ratios, in particular [O~\textsc{i}]/[Ca~\textsc{ii}]~$\approx$~0.71, are consistent with progenitor mass estimates of 12--20~M$_\odot$. This, combined with the relatively low ejecta mass and rapid light-curve decline ($\Delta m_{15}(V) \approx 1.03$), supports the interpretation of SN~2015ap as arising from a moderate-mass star in a close binary system rather than from a single massive Wolf--Rayet star. The -- approximately -- solar metallicity in the explosion site further strengthens this picture. These findings set the stage for a more targeted investigation of binarity in stripped-envelope SNe~Ib. 

In this work, we present a comprehensive re-analysis of SN~2015ap, combining detailed multi-band photometry and spectroscopy with advanced modeling techniques to robustly test and further constrain the binary progenitor scenario.

\begin{figure*}
  \centering
    \includegraphics[scale=0.5]{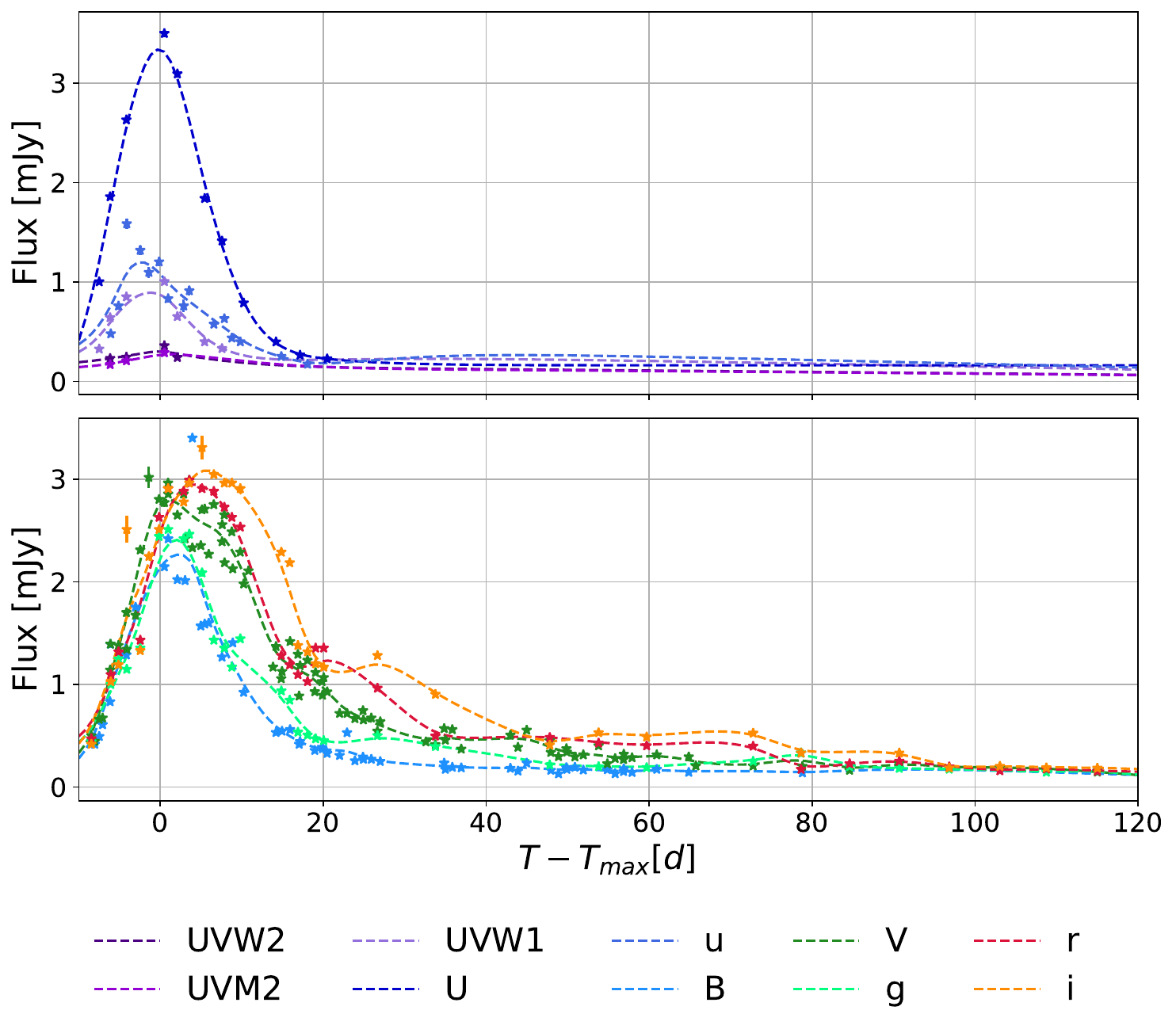}
  \caption{Multi-wavelength light curves of SN 2015ap. For further details on the analysis, see Sect.~\ref{LS_analysis}. The dashed lines represent the Gaussian Process fits to the light curves in each filter. The time axis is given relative to the epoch of the V-band maximum, $T_{\rm max}$.}
  \label{fig:LC-SN2015ap}
\end{figure*}

\section{Data}

SN~2015ap was discovered on 2015 September 8 by the 0.76-m Katzman Automatic Imaging Telescope \citep[KAIT;][]{2000AIPC..522..103L} as part of the Lick Observatory Supernova Search (LOSS; \citealt{flipper2001}) in an unfiltered 18~s image \citep{ross2015supernova}. Its host galaxy IC 1776 has a redshift of z = $0.011375 \pm 0.000017$ \citep{1993Chengalur} and a barred spiral morphology.
The SN was classified as a type Ib a few days after maximum brightness on the basis of well-developed features of He I, Fe II, and Ca II \citep{2020Gangopadhyay}.

In this work, we make use of publicly available photometric and spectroscopic data of SN~2015ap presented by \citep{Prentice_2018}, \citet{2020Gangopadhyay} and \citet{Aryan2021}, obtained through extensive follow-up campaigns using multiple ground-based telescopes (e.g., HCT, Copernico, DFOT, NOT, KAIT). The dataset includes multi-band photometry in Johnson-Cousins \textit{UBVRI}, SDSS \textit{ugri}, as well as ultraviolet measurements in \textit{UVW2}, \textit{UVM2}, and \textit{UVW1} taken with the \textit{Swift}/UVOT instrument \citep{2005SSRv..120...95R, 2014Ap&SS.354...89B}\footnote{Swift's Optical/Ultraviolet Supernova Archive (SOUSA); \url{https://archive.stsci.edu/prepds/sousa/}}. Altogether, the photometric coverage spans from the far ultraviolet (approximately 100–200 nm) to the red edge of the optical regime (around 700 nm), approaching the near-infrared region (starting near 700 nm and extending up to about 2500 nm).

Low- and medium-resolution optical spectra of SN~2015ap were acquired between phases $-7$ and $+235$ days relative to $V$-band maximum. These spectra were obtained with the 2.0m Himalayan Chandra Telescope (HCT) and the 1.82m Copernico Telescope in Asiago, and cover the wavelength range from $\sim3400$~\AA\ to $\sim8700$~\AA.

The full photometric and spectroscopic dataset was already reduced and calibrated as described in \citet{2020Gangopadhyay},
We adopt $A_{V,host} = 0.174 \pm 0.198$ mag for SN 2015ap, as estimated in \citet{2023ApJ...955...71R} based on NaID equivalent width and color curves. Although this value is small and consistent with zero within $1\sigma$, it is slightly higher than the Milky Way extinction ($A_{V,MW} = 0.115$ mag, also reported in \citet{ned2021}), suggesting that host extinction may not be negligible. We adopt redshift $z = 0.01138$, and distance modulus $\mu = 33.27 \pm 0.15$~mag throughout this work.

In the present study, we reanalyse these data with the aim of constraining the binary nature of the progenitor, focusing on detailed modelling of the light-curve evolution and nebular spectra beyond what was performed in the original analyses.

\section{Photometric analysis}\label{LS_analysis}
\begin{figure}    
    \includegraphics[scale=0.3]{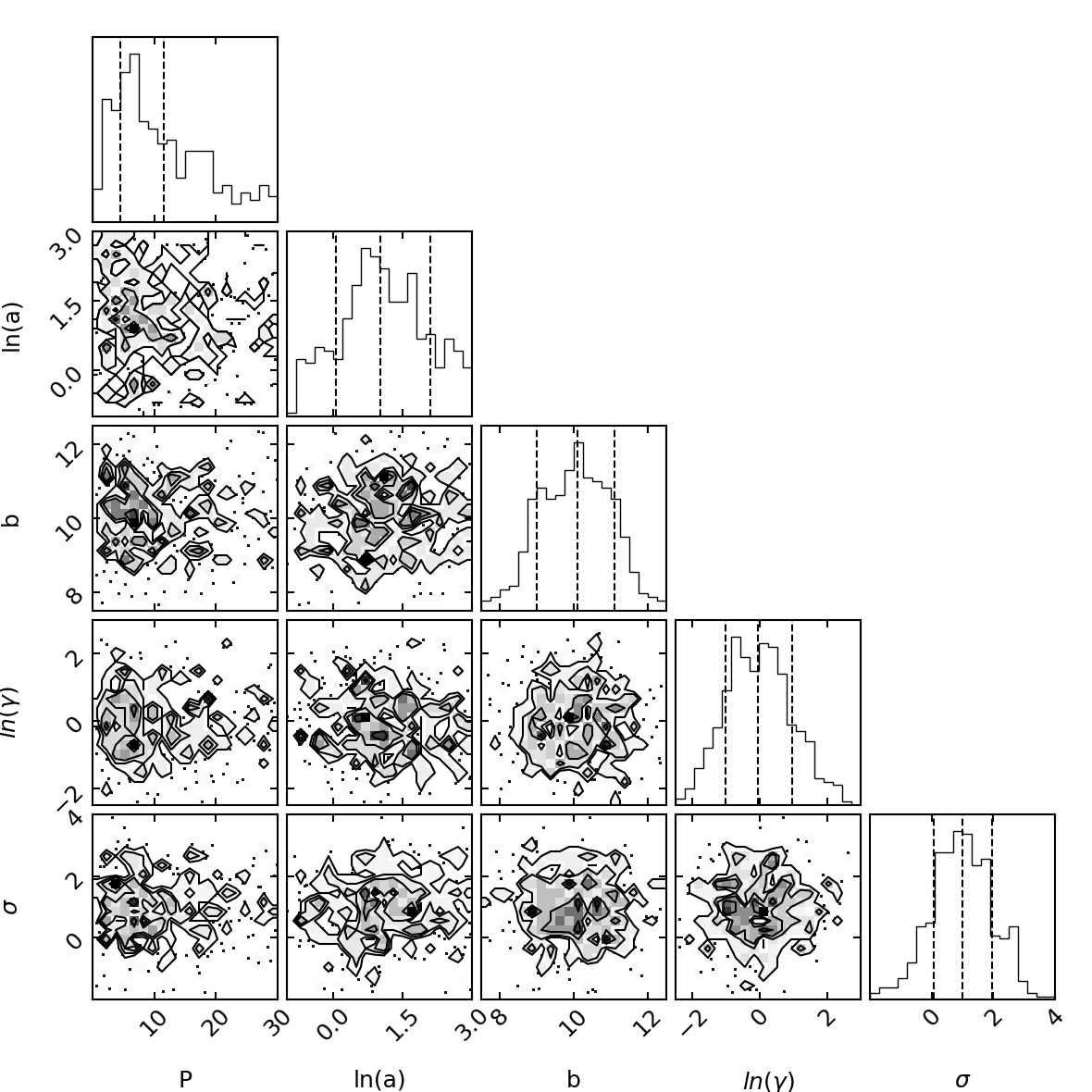}\quad \quad
     \includegraphics[scale=0.45]{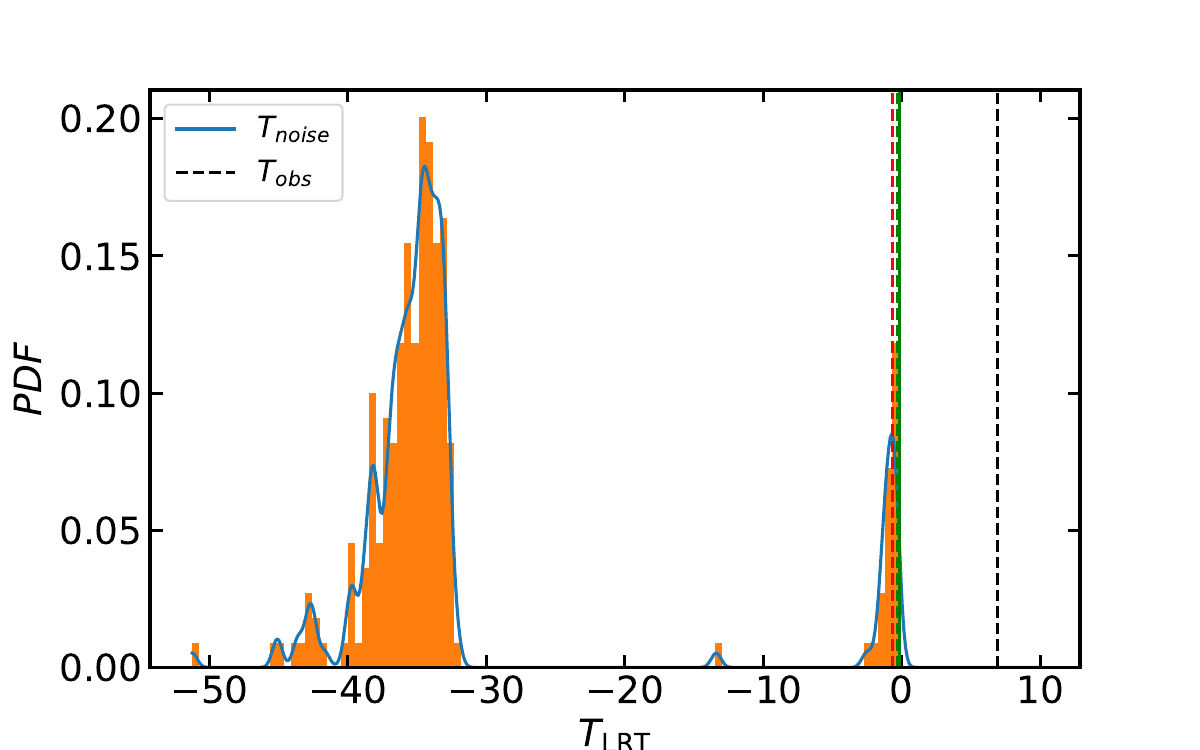}
    \caption{The figure shows the example for the B band of the periodic analysis via \texttt{$mind\_the\_gaps$}. Top panel: the corner plot for the fitting of the hyperparameters of the GP Kernel; bottom panel: the distribution of the likelihood ratio test of the periodic model with respect random fluctuation model. }\label{fig:pulseB}
    
\end{figure}
\begin{figure*}
  \centering
    \includegraphics[scale=0.3]{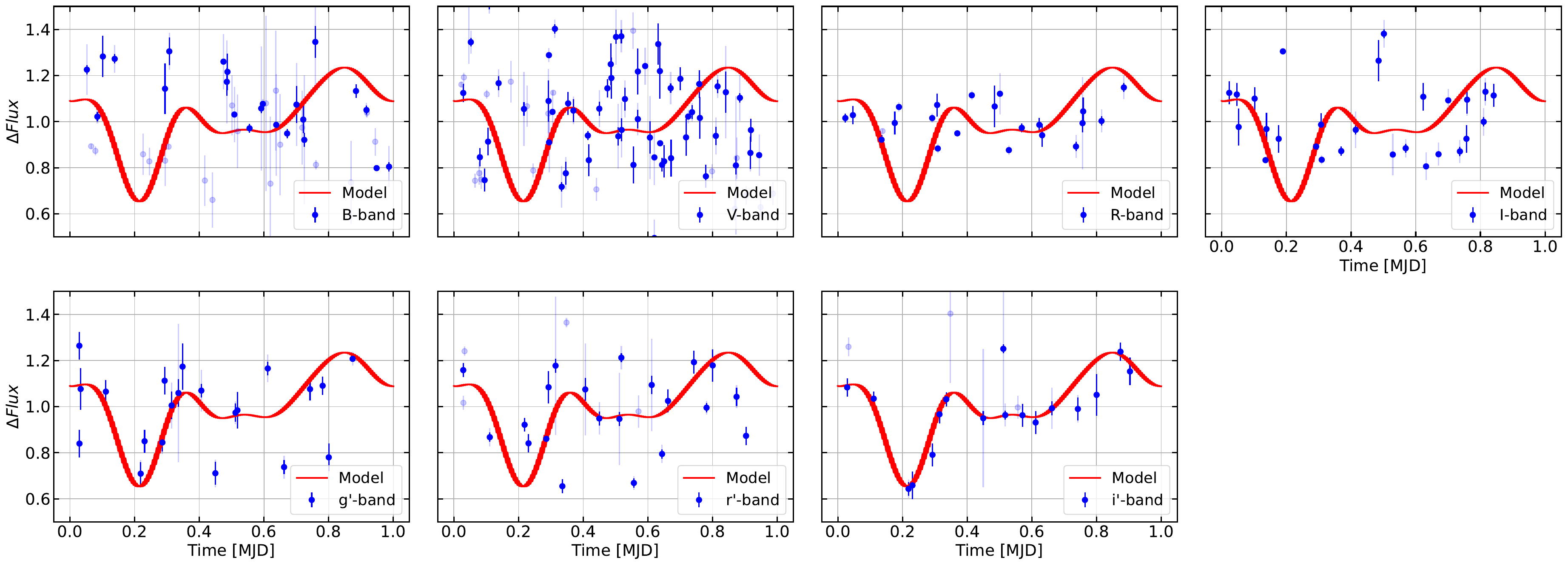}
  \caption{Phase folded light curve comparison of the inferred evolution of the bumps as predicted by the GP model (red bold line) and the observed evolution in different bands. The shaded dots are the observations with a signal-to-noise ratio, SNR, less than 10. }
  \label{fig:period_lc}
\end{figure*}
\begin{figure}    
    \includegraphics[scale=0.35]{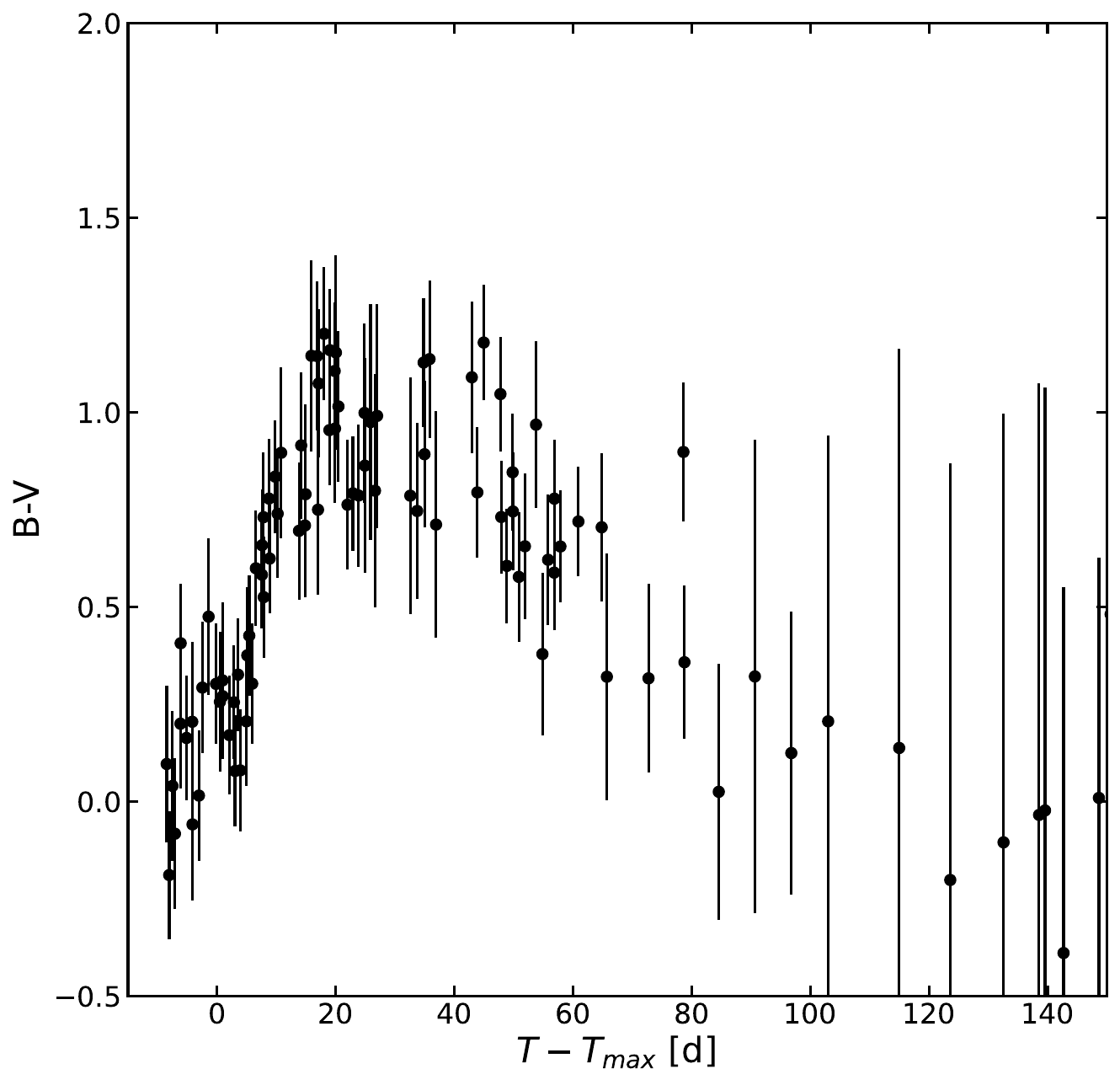}
    \caption{B-V colour evolution of SN~2015ap. The chromatic and thermal behavior suggests a prolonged breakout or delayed ignition of energy.}\label{fig:color_BV}
    
\end{figure}
\begin{figure*}
  \centering

    \includegraphics[scale=0.3]{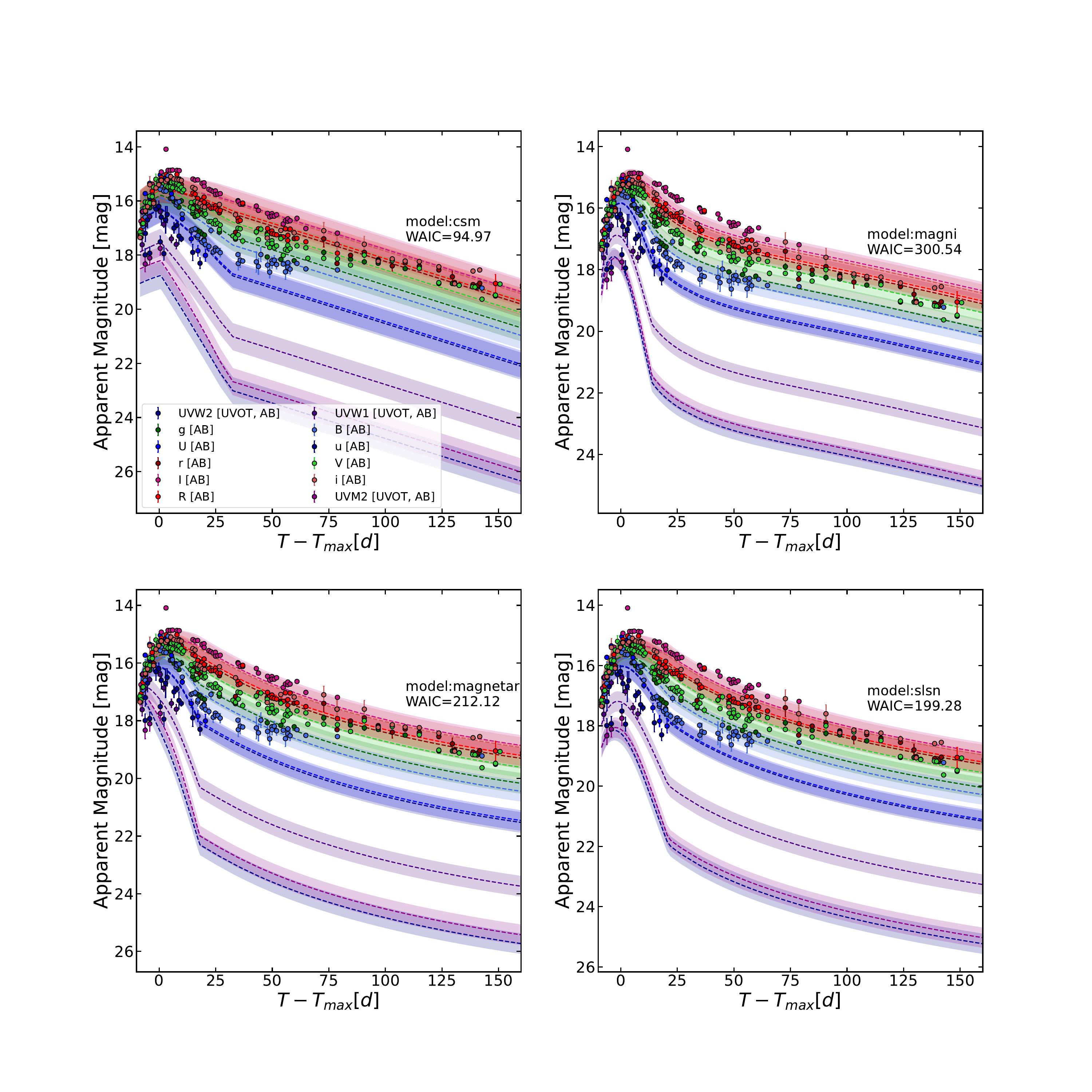}
  \caption{Light curve modeling for SN~2015ap with different power source scenarios as described in \autoref{tab:modeltable}. For more details on the analysis, see Sect. \ref{modelling}.{The points in each panel are the observations of the SN in different bands. The dashed lines follow the color scheme indicated in the legend and represent the best-fit realization of each model. The shaded regions correspond to the 1$\sigma$ uncertainty intervals. Each panel shows a different model, for which the WAIC is reported as a measure of the model’s ability to replicate the data.}}
  \label{fig:models_lc}
\end{figure*}

\begin{figure*}
\hspace{-1cm}
  \includegraphics[scale=0.35]{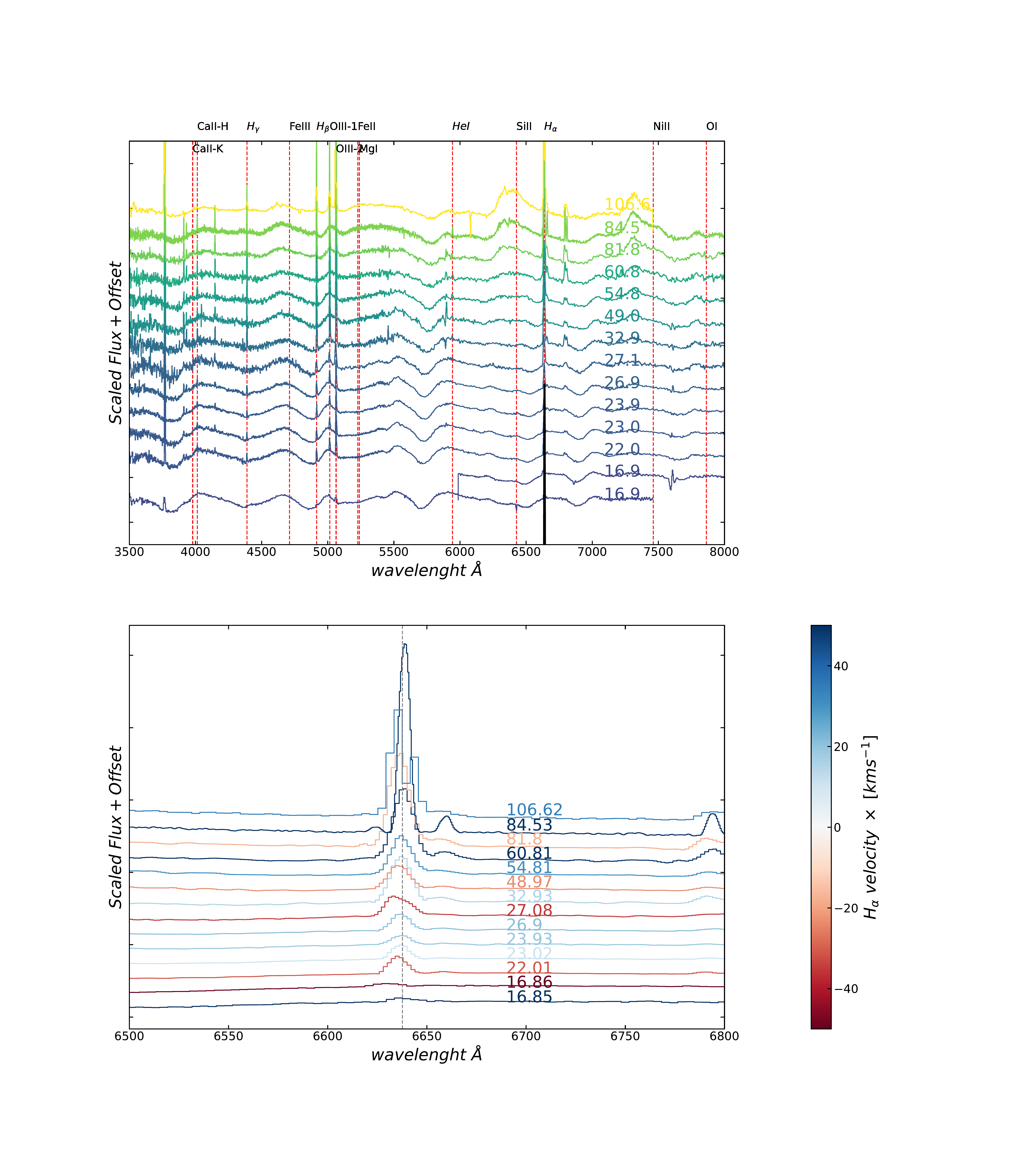}
  \vspace{-1cm}\caption{Multi-epochs spectra of SN~2015ap (top panel) color coded with respect to the temporal evolution of the spectra in MJD; and a zoom of the $H_{\alpha}$ line color coded with its velocity (bottom panel). It appears that the peak line velocities vary periodically through the entire evolution of the spectra (bottom panel). The labels of the phase that appear on the right side of all the spectra are estimated with respect to the very first available photometric observation.}
  
  \label{fig:spectra}

\end{figure*}

\begin{figure}
  \centering

    \includegraphics[scale=0.4]{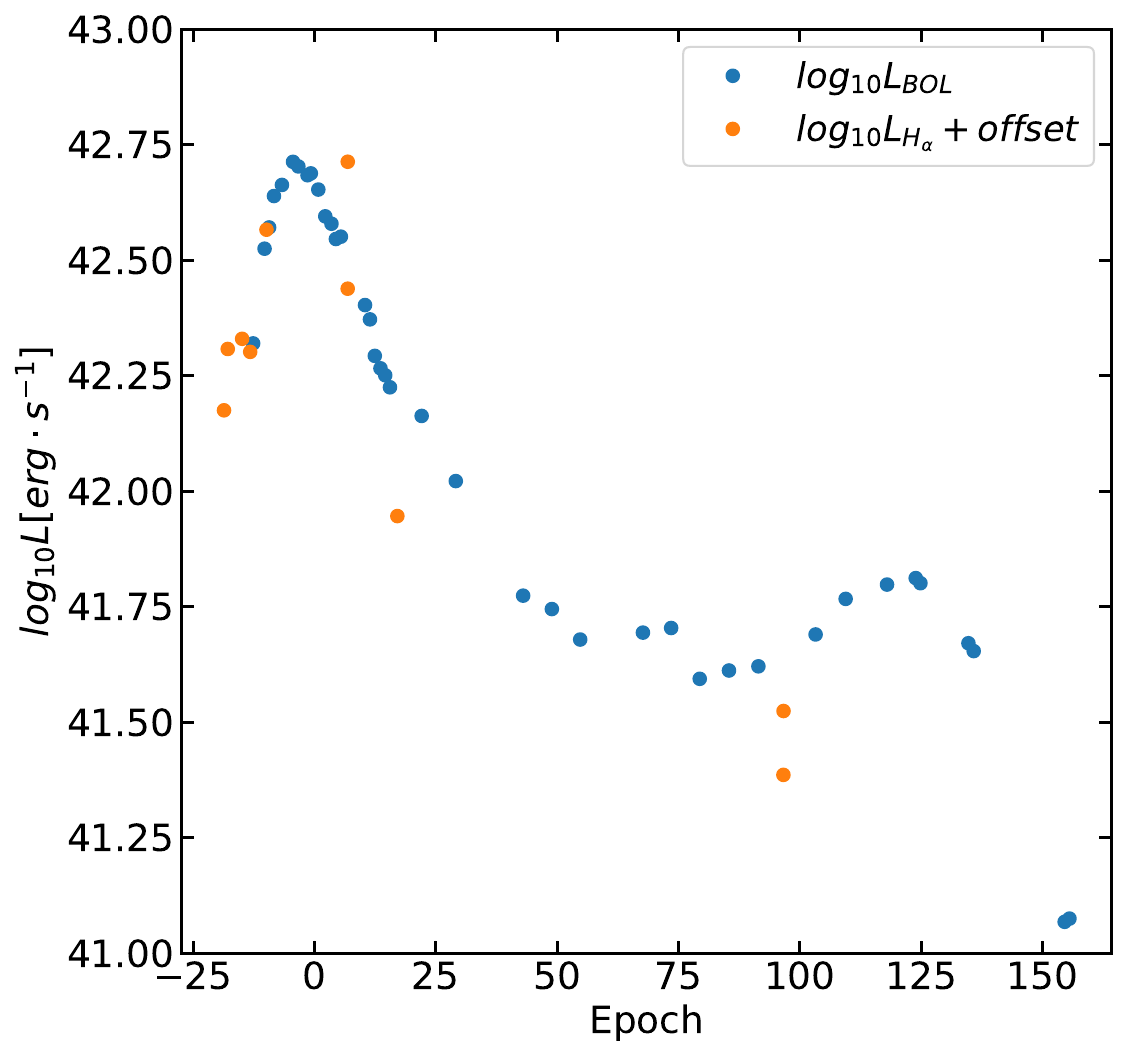}
  \caption{Evolution of the $H_{\alpha}$ luminosity with the bolometric light curve of SN~2015ap. The close correspondence between the two suggests that both emissions originate from the same region within the ejecta.}
  \label{fig:hlum}
\end{figure}

\begin{figure}
\centering
    \includegraphics[scale=0.4]{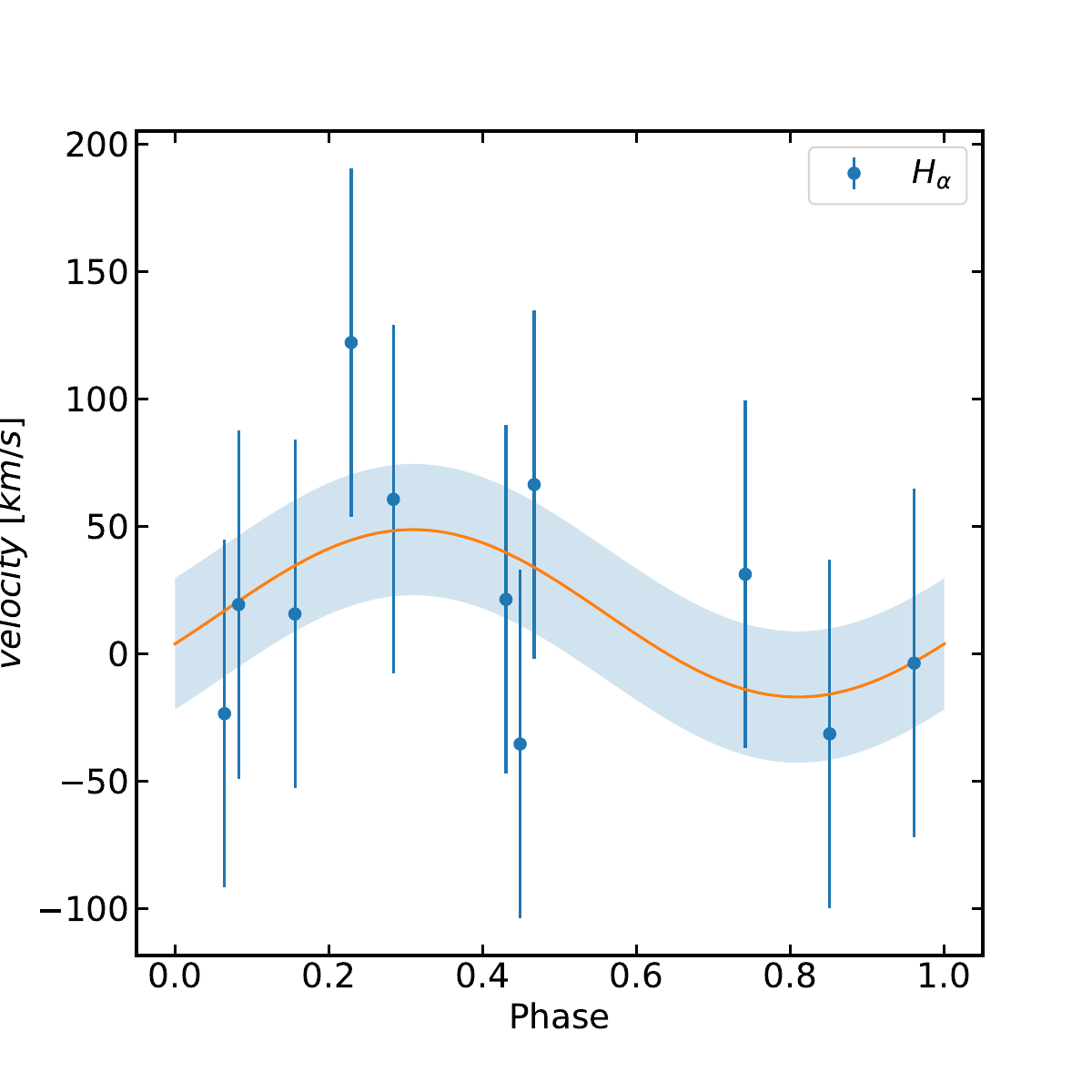}
    \caption{The phase folded velocity curve of $H_{\alpha}$ line.{The figure shows the fit of the velocity curve with a sinusoid, assuming the period derived with the Gaussian Process. The shaded area defines the model's $3-\sigma$ uncertainty region.} 
    }
    \label{vel_model}
\end{figure}

\begin{figure*}
  \centering

    \includegraphics[scale=0.4]{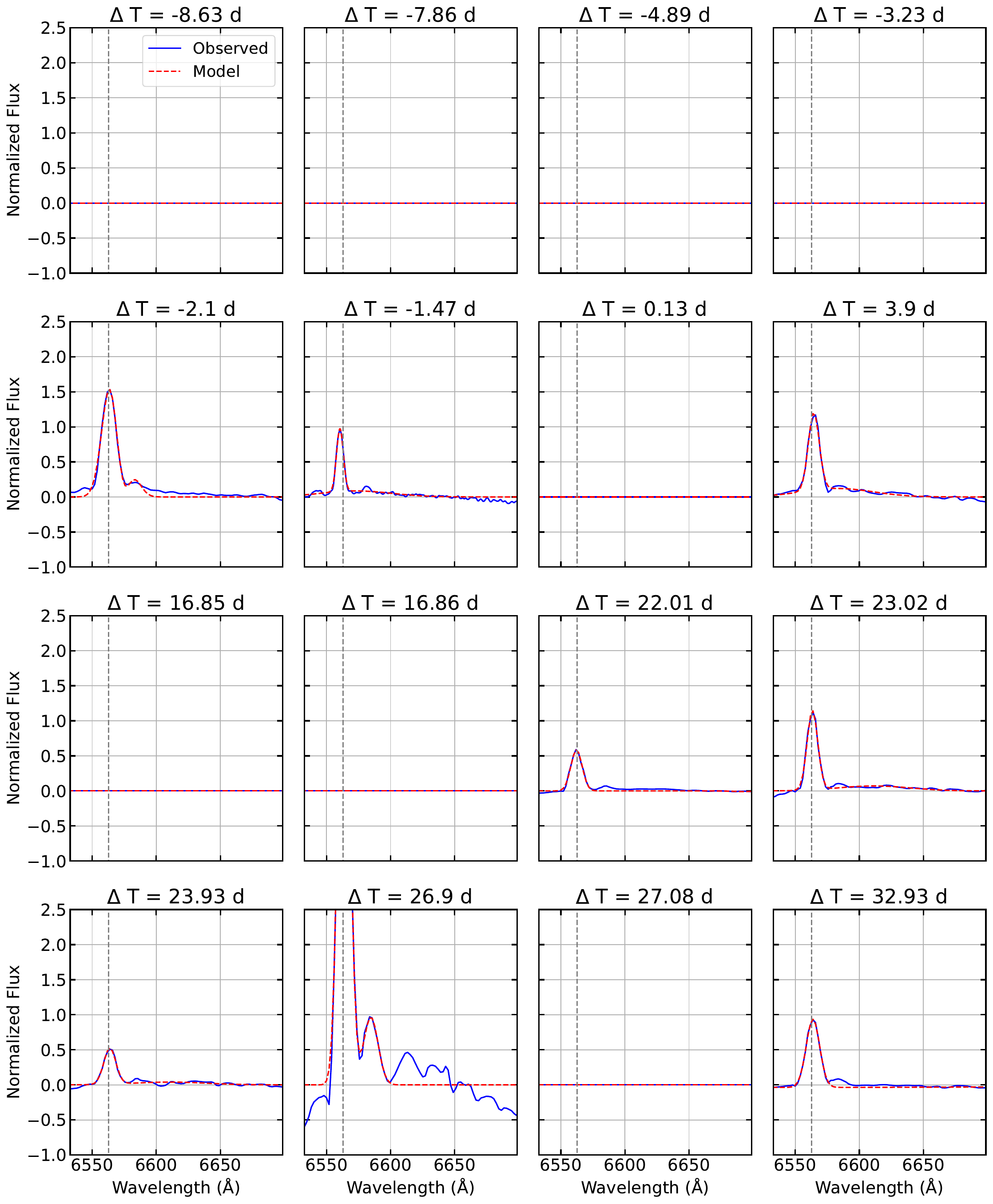}
  \caption{The figure displays the evolution of the observed $H_{\alpha}$ line across multiple epochs, in comparison with the predicted emission from a shell-distributed CSM model (shown as a red dashed line).}
  \label{fig:model_h}
\end{figure*}

\begin{figure}   
    \includegraphics[scale=0.35]{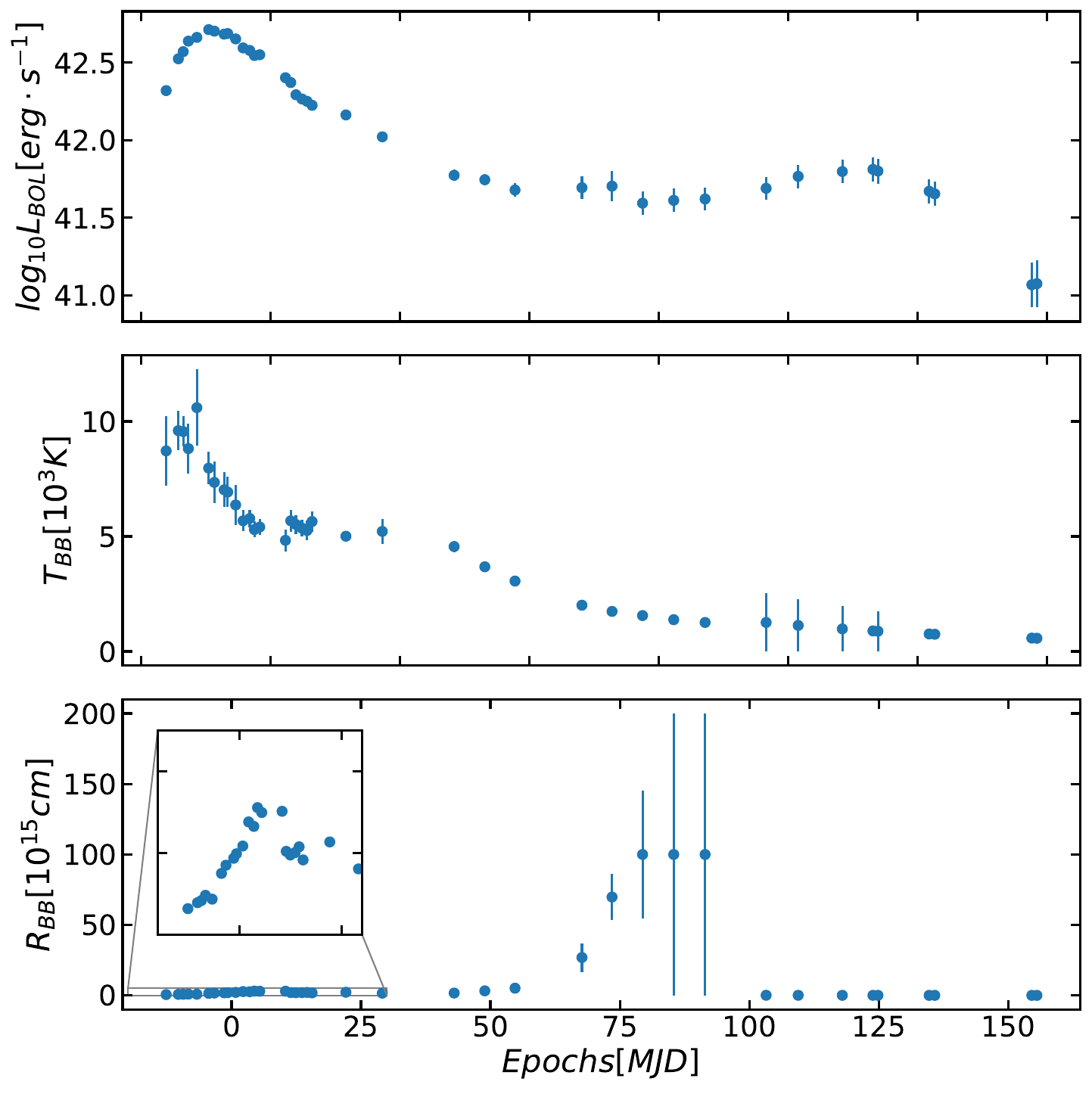}
   \caption{Results of bolometric light curve estimation using \texttt{superbol}. Top panel: bolometric light curve including the additional black body corrections. Middle panel: evolution of the estimated black body temperature with time from the maximum of the bolometric light curve. Bottom panel: evolution of the estimated black body radius with time from the maximum of the bolometric light curve. } \label{fig:superbol}
    
\end{figure}

SN~2015ap appears as a typical type-Ib SN. A closer inspection of the decay tail of the light curves (Fig. \ref{fig:LC-SN2015ap}) reveals the presence of photometric fluctuations {in the redder bands on the time scale of $10$ d. These are visible with similar amplitude in V, g, r and i bands} and across different telescope and instrument combinations, indicating that this is neither an instrumental nor a calibration effect. 

To identify and quantify potential periodicities, we first subtracted the decline trend from the light curve. To separate the contribution of the decline due to $^{56}$Ni from the modulation, we applied a wavelet decomposition, which has been performed using the Python \texttt{pywt} package 
\citep{pywt}.
Hence, we extrapolated the lowest-order component that describes the large-scale structure of the light curve. The modulation contribution in the photometric evolution comes from subtracting the large-scale structure from the light curve. 

A Lomb-Scargle analysis would generally be better suited for highlighting the presence of periodicity in a time series. However, the limited data on the light curve's decay tail, combined with the heteroscedastic distribution of the data over time, during the later stages of the SN evolution, introduces bias in the analysis. This is because a proper definition of the Nyquist frequency is impossible under these conditions \citep{1996Israel, 2005Vaughan}.
Thus, we performed an autocorrelation analysis to highlight the presence of a periodic behaviour in the data.{The ACF was computed using an interpolation-based discrete autocorrelation method, which first reconstructs the light curve on a regular grid through interpolation, and then calculates the autocorrelation at evenly spaced lags. This approach mitigates the effects of uneven sampling and allows for a more robust identification of periodic signals.}
In the following, we discuss the case for the B band as representative of the results for all the other bands. The analysis of the Autocorrelation Function (ACF) has revealed the presence of a significant periodicity of $6.0 \pm 0.8$ days, with a probability of $95\%$ to find the measured value within the uncertainty region. This periodicity corresponds to the most prominent peak of the ACF, indicating a recurring variability in the source.
{To verify the consistency across filters, we repeated the same analysis for each band and compared the resulting periods. All estimates are compatible within 1$\sigma$ uncertainties, and the periodicity remains stable despite the different cadences and sampling patterns.} 
Even though the ACF peak just surpasses the $3\sigma$ limit, the coherence of the signal across all bands supports the hypothesis of a genuine periodic modulation.
{The ACF, however, implicitly assumes a virtually homogeneous cadence, which is not present in the data and cannot be reconstructed. As a result, the outcome of the ACF analysis cannot be considered fully reliable in the presence of irregular sampling.}

Thus, we adopted a Gaussian Process-based approach as described in \citet{Gurpide2025}\footnote{Code available at \url{https://github.com/andresgur/mind_the_gaps/tree/main}.}, named \texttt{$mind\_the\_gaps$}, which provides a flexible probabilistic framework for modeling time series data while explicitly accounting for temporal correlations. Unlike Fourier-based methods, GP modeling operates directly in the time domain, ensuring a well-defined likelihood function that can be leveraged for hypothesis testing. We considered this approach with respect to standard ones because the latter tend to overestimate the significance of periodic signals, increasing the likelihood of false positives, when stochastic variability (e.g., red noise) is present.

Hence, we considered a Bayesian fitting procedure to deeply scan the parameter space of a periodic-based Gaussian Process. We extracted periodicity measurement from the posterior distribution of the parameter space of a periodic kernel 
$$
k(x_i, x_j) = \text{exp}\left(-
\frac{ 2\sin^2(\pi d(x_i, x_j)/P) }{ \gamma^ 2} \right).
$$
where $\gamma$ is the length scale of the kernel, $p$ is the periodicity of the kernel and $d(\cdot,\cdot)$ is the Euclidean distance.

We populated the posterior through the Monte Carlo Markov Chain method \citep[for details see, e.g.,][and references therein]{2019arXiv190912313S}, implemented in the \texttt{emcee} Python package \citep{2013PASP..125..306F}.
{Across all filters, the period estimates are consistent within their respective uncertainties. By computing a weighted mean, we find that the period corresponding to the peak of the distribution is $P\approx 8.41\pm 0.80$ d (see top panel in Fig. \ref{fig:pulseB}).}

By constructing likelihood ratio tests (LRT) using posterior distributions derived from the inferred noise properties, we assess the significance of candidate periodic signals while minimising the risk of misidentification (see bottom panel in Fig.~\ref{fig:pulseB}). We inferred the improvement in adding the periodic model simulating $1000$ time series from the posterior distribution of the noise and measuring the LRT with respect to the periodic model.
The observed $T_{LRT}$ is $6.9$, which falls in the outer region of the distribution, indicating that adding the periodic signal significantly improves the model, suggesting that the periodicity is real.

The comparison between the best fit and the data is presented in Fig. \ref{fig:period_lc}. The paucity of data is evident, and this could have caused a spurious detection of a periodicity due to noise. 

The observed bumps analysed here are also reflected in the colour evolution (Fig.~\ref{fig:color_BV}), which shows an initial blueward trend corresponding to the rising phase of the photometric evolution. The observed trend then reverses for  $\sim$10 days, after which a steep shift back toward the blue part of the spectrum is marked by a second peak centred at $T-T_{max} \simeq20$ d, suggesting a supplementary energy injection source at that epoch. Typically, this second peak is not expected to be so pronounced if the temperature increase is interpreted as due to typical  CMS interaction, from CSM originating from the progenitor's steady wind. 
Eventually, the rapid temperature variation and the non-linear evolution of the B-V colour curve suggest the presence of a prolonged brakeout phase. 

A similar effect has been observed in iPTF13bvn \citep{2016Fremling}, supporting the idea that some progenitor stars are in a binary.
The trend observed in the colour evolution of SN 2015ap has been observed in other SNe, such as SN 1999ex, SN 2008D, and iPTF13bvn \citep{2016Fremling, 2019Yoon, yoon2010type, 2015AMoriya, 2002Stritzinger, 2009Modjaz}, and the differences in mixing suggest a different origin of the progenitors. In particular, \citet{2019Yoon} suggests that the progenitors of SN Ib and SN Ic differ in terms of He content and mixing of the $^{56}Ni$.
The model with a binary progenitor is compatible with weak mixing of the $^{56}Ni$, since mass transfer can remove part of the outer layers without strong mixing of the radioactive elements in the upper layers of the envelope.
\subsection{Light curve modelling}\label{modelling}

\begin{table*}
    \centering
    \begin{tabular}{|c|c|c|}
        \hline
        Model name & Descriptions & References \\
        \hline
        csm    & Interacting CSM-SNe    & \citet{Chatzopoulos}    \\
        \hline
        magnetar    & Magnetar engine with simple SED    & \citet{Nicholl}    \\
        \hline
        magni    & Magnetar engine with simple SED + NiCo decay   & \citet{Nicholl,Nadyozhin}    \\
        \hline
        slsn    & Magnetar + modified SED + constraints   & \citet{Nicholl}    \\
        \hline
    \end{tabular}
    \caption{Description of the \texttt{MOSFiT} models used for the model fitting}
    \label{tab:modeltable}
\end{table*}

In \citet{Prentice_2018} and \citet{Aryan2021} SN~2015ap is described as a He-rich SN with a photospheric velocity near luminosity peak $\approx$9000 km s$^{-1}$. With the known values of the photospheric velocity near maximum light, $t_\mathrm{rise}$, and a constant opacity $\kappa$ = 0.07 cm$^2$ g$^{-1}$, \citet{Aryan2021} also obtain an ejecta mass $M_\mathrm{ej}=2.2 \pm 0.6 \, \mathrm{M_\odot}$ and kinetic energy $E_k= (1.05 \pm 0.31) \times 10^{51}$ erg from the \citet{1982Arnett} model with a $M_{^{56}Ni} = 0.14\pm0.02 \, \mathrm{M_\odot}$.

We used simple light curve modeling to estimate the ejecta mass of SN~2015ap using the Modular Open Source Fitter for Transients \citep[MOSFiT;][]{Guillochon_2018}. MOSFiT is a publicly available code that we employed to fit semianalytic models to the multiband observed light curves of SN~2015ap. All model fitting was performed using the dynamic nested sampler DYNESTY package \citep{2020Speagle} option in MOSFiT.
In our analysis, we fitted the light curves using four physically motivated models, each representing a distinct powering mechanism (see table \ref{tab:modeltable} for a summary). The first is the csm model, which assumes that the luminosity arises from the interaction between the supernova ejecta and a dense circumstellar medium (CSM). This model accounts for both forward and reverse shock contributions and is particularly suitable for hydrogen-rich superluminous supernovae, as described by \citep{Chatzopoulos_2013}. The second model, magnetar, is based on energy input from the spin-down of a newly born magnetar, assuming a simple blackbody spectral energy distribution (SED) with absorption. This framework was introduced by \citep{Nicholl} and implemented in the MOSFiT fitting tool. The magni model extends the basic magnetar scenario by including additional power from the radioactive decay of $^{56}$Ni to $^{56}$Co, following the prescriptions of \citep{Nadyozhin} and \citep{Nicholl}, thus improving the fit to the late-time tail of the light curve. Finally, the slsn model builds on the magnetar framework but introduces a modified SED with suppressed ultraviolet emission below $3000\AA$, time-dependent photospheric evolution, and empirical constraints specific to hydrogen-poor superluminous supernovae (SLSNe-I), making it the most physically detailed model among those considered.
Fig. \ref{fig:models_lc} and table \ref{tab:MOSFiT_unbound} show the results from the fitting procedure assuming all parameters are unbound.

\begin{table*}
    \centering
    \begin{tabular}{|c|c|c|c|c|c|c|c|c|}\hline
        model name&$d_L$& $B $& $M_{NS}$ &$P_{spin}$&$M_{ejecta}$ &$v_{ej}$& $M_{CSM} $ & $f_{^{56}Ni}$  \\
         &$\mathrm{Mpc}$& $10^{14} \, \mathrm{G}$& $\mathrm{M_\odot}$ &$10^{-3} \, \mathrm{s}$&$\mathrm{M_\odot}$ &$\mathrm{km \, s^{-1}}$& $\mathrm{M_\odot}$ &  \\ \hline
         csm&$21.88$ & -&-&-&0.94&1995&0.83 & - \\ \hline
         magnetar&-&0.15&1.77&1.23&0.001&14791&-&-\\ \hline
         magni&-&6.92&1.21&5.2&0.67&12022&-&0.44 \\ \hline
         slsn&79.43&3.9&1.01&9.90&1.26&12104&-&-  \\ \hline
    \end{tabular}
    \caption{Model parameters from MOSFiT. Uniform distributions are used as priors for all the parameters. Uncertainties on the parameters are the $10\%$ of the best-fit values.}
    \label{tab:MOSFiT_unbound}
\end{table*}

In \citet{2019Prentice} and \citet{Aryan2021, 2022JApA...43...87A}, SN~2015ap is described as a He-rich type Ib SN. Compared to their best fit values, our modelling with MOSFiT yields significantly lower ejecta masses for some models (e.g., $M_\mathrm{ej} \approx 0.94$ M$\odot$ in the CSM model) and higher kinetic energies when implied by the velocities (e.g., $v_\mathrm{ej} \approx 12000$–$15000$ km s$^{-1}$ for most models). It is also interesting that with respect to the magnetar model, the inferred parameters show a good agreement for the magnetic field intensity, but lower spin period. Notably, our modeling incorporates photometry from additional UV bands provided by the Swift/UVOT filters, which were not included in either of the earlier analyses. These UV data significantly improve the constraints on the early rising part of the light curve, which is particularly sensitive to the heating mechanism and the presence of additional energy sources like a magnetar or CSM interaction.

Eventually, the inclusion of UV data appears to shift the preferred model towards CSM interaction, as suggested by the Watanabe-Akaike Information Criterion (WAIC), analysis, although the rise and tail shapes are not perfectly reproduced. This mismatch could be due to the simplified assumptions in the CSM density profile within MOSFiT or to a composite powering mechanism involving both interaction and central engine input. The larger velocity inferred in our models could also reflect the fact that UV observations are more sensitive to hotter, faster-moving outer layers, further affecting the inferred ejecta structure.

\section{Spectra analysis}\label{sec:spectra}

The spectra reveal prominent features of He I, Mg II, Fe II, Si II, and O I, consistent with the properties of type Ib SNe. The He I 5876 \AA\ line shows an expansion velocity of approximately $15,500$ km s$^{-1}$, characteristic of stripped-envelope SNe.

The galaxy-subtracted spectral evolution of SN~2015ap is presented in the top panel of Fig. \ref{fig:spectra}.
We extracted the line intensity by subtracting the pseudo-continuum, which includes contributions from unresolved absorption/emission features, scattered light, and possible instrumental artefacts. The continuum model is constructed by fitting a linear model to the arbitrarily selected continuum region on both sides of the emission feature. We measured the line velocities using the flux-weighted centroid of the emission feature, without modelling its complex velocity structure with a Gaussian fit. The velocity uncertainties include contributions from flux errors and variations due to different choices of the continuum region. 
We measured the velocities for all the isolated emission lines in the evolution of multi-epoch spectra. We avoided the lines that showed the P-Cygni profile which inherited a complex structure that bias the measure.
We used \texttt{specutils}\footnote{\url{https://doi.org/10.5281/zenodo.10681408}} package to analyze spectral data. 
To verify that we properly excluded galaxy contribution, we measured the $H_{\alpha}$ luminosity evolution with time. We were mainly interested in the $H_{\alpha}$ line, due to its link with accretion phenomena, which in our case can indicate the presence of a companion.
$H_{\alpha}$ luminosity evolution matched the trend of SN~2015ap's bolometric light curve evolution (see Fig. \ref{fig:hlum}), indicating that the emission we are measuring comes from the region of the ejecta itself.

The early-time spectra show a blue continuum, which is characteristic of high-energy events like SN explosions. This blue shift indicates that the ejecta are moving away from the observer at high velocities.
Zooming around the iron absorption line (Fig. 16 in \citealt{2020Gangopadhyay}), it is visible a 'W' Shape Absorption Feature: the spectra taken at 5 and 7 days post-explosion reveal a 'W'-shaped absorption feature around 4000 \AA, which is associated with Fe complexes.

The spectral features, including the 'W' shape, suggest that the progenitor of SN 2015ap is likely a star with a mass between 12 and 20 $\mathrm{M_\odot}$. This estimation is supported by the analysis of the [O I]/[Ca II] ratio and nebular spectral modelling, which indicate a low-mass star in a binary system \citep{Kuncarayakti}. The [O I]/[Ca II] ratio is a significant diagnostic tool in determining the progenitor mass. In the case of SN 2015ap, this ratio is approximately 0.71, which is consistent with low-mass progenitors in binary systems. This ratio is influenced by various factors, including temperature and density, and serves as a demarcation between binary and single progenitor systems.
{We use} the relation:
$$
M_{O} = 10^8f([\mathrm{OI}])D^2e^{\frac{2.28}{T_4}}
$$

where $M_O$ is the mass of the neutral O in $\mathrm{M_\odot}$ units, D is the distance to the galaxy in Mpc, $f([\mathrm{OI}])$ is the total flux of the [OI] $6300,~ 6364~ \AA$ feature in $\mathrm{erg s^{-1} cm^{-2}}$, and  $T_4$ is the temperature of the O-emitting region in units of $10^4$ K \citep{1986ApJ...310L..35U}.  Using the observed flux of $4.0 \times 10^{-15} \, \mathrm{erg s^{-1} cm^{-2}}$ of the [OI] $6300, 6364 \AA$ doublet from $57389.08$ MJD spectrum, and adopting $T_4 = 0.37 ~$K, we estimate $M_O \approx 0.47 \, \mathrm{M_\odot}$. With the assumption that at most the $50\%$ contribute to the emission, we estimate that the total amount of oxygen should not exceed $0.9 \, \mathrm{M_\odot}$.
The nebular spectral modelling in \citet{2020Gangopadhyay} indicates that the progenitor is most likely a star in the range of 12 to 17 $\mathrm{M_\odot}$, with a strong likelihood of being in a binary association. The modelling takes into account the observed line luminosities and the overall spectral characteristics.

The indication that SN 2015ap's progenitor is a high-mass star in a binary system has important implications for its evolution. In binary systems, mass transfer can occur, affecting the final mass and composition of the progenitor star before it undergoes a SN explosion. This interaction can lead to the stripping of outer layers, resulting in the stripped-envelope nature of type Ib SNe like SN 2015ap \citep{yoon2010massive, Eldridge2013, 2016Liu, Solar:2024rar}.
SN~2015ap shows an interesting feature of $H_{\alpha}$, characterised by a very narrow feature on top of a broader one. The wavelength range of interest in the SN spectrum could be contaminated by host galaxy emission from nearby H II regions. However, checking the spectra of the host at different epochs, we confirm that the narrow feature does not come from any artefact due to the host galaxy line contamination.
As one can see from the bottom panel in Fig.\ref{fig:spectra}, the narrow feature  wavelength oscillates around the $H_{\alpha}$ rest-frame wavelenght. {By fitting a sinusoidal model to the phase-folded velocity curve using the period derived from the SN light curve analysis, we found that it accurately reproduces the observed trend. This confirms that the velocity shifts exhibit a cyclical pattern consistent with the 8.4-day modulation detected in the optical light curves  (see Fig. \ref{vel_model})}

Similarly to SN 2022jli \citep{2023Moore,2023Chen}, we interpret the periodic undulation in the SN~2015ap light curve coming from a limited region with relatively small size in the centre of the ejecta. This implies, similarly, the emission of $H_{\alpha}$ also comes from the centre of the ejecta. The hydrogen responsible for the $H_{\alpha}$ emission likely originates from the binary interaction in the previous phases of the explosion. 
$H_{\alpha}$ emission lines have been commonly observed in binary systems with accretion disks, from Roche-Lobe interaction \citep{2019Zamanov, 2025Zhuang}, which could serve as an analogy in the low accretion rate regime to understand the emission mechanism and structure of the $H_{\alpha}$ in SN~2015ap as in SN~2022jli. This would be attributable to the presence of a shell-distributed CSM. 
To analyse the temporal evolution of the CSM around SN 2015ap, we modelled the observed H$_{\alpha}$ line profiles as a superposition of Gaussian absorption/emission features originating from multiple expanding shells. This approach enables a simplified yet physically motivated representation of the kinematics and optical depth of the CSM.

Each shell is characterised by four main parameters: its expansion velocity $v$, the velocity dispersion (or thermal broadening) $\sigma$, the optical depth $\tau$, and a velocity shift $\Delta v$ to account for asymmetries or additional line-of-sight velocity components. The total synthetic line profile is then given by the sum of Gaussian components shifted according to the relativistic Doppler formula:
\[
\Delta \lambda = \lambda_0 \frac{v + \Delta v}{c},
\]
where $\lambda_0$ is the rest-frame wavelength of the transition and $c$ is the speed of light.

The full model flux profile is:
\[
F(\lambda) = \sum_{i} \tau_i \exp\left[ -\frac{1}{2} \left( \frac{\lambda - (\lambda_0 + \Delta \lambda_i)}{\sigma_i} \right)^2 \right],
\]
with $\sigma_i$ related to the shell's velocity width as $\sigma_i = \lambda_0 \frac{\text{width}_i}{c}$.

This model was implemented in Python using the \texttt{astropy} library for physical unit handling and the \texttt{scipy.optimize.curve\_fit} function to fit the model to the continuum-subtracted spectra at different epochs. The continuum was estimated and subtracted using a spline-based method from \texttt{specutils.fit\_generic\_continuum}. Similar approaches have been used in studies of type IIn SNe and other CSM-interacting transients \citep[e.g.,][]{2017Dessart, smith2017observational}.
An example of the shell model result can be seen in Fig. \ref{fig:model_h}.

The velocity evolution observed in the $H_{\alpha}$ line profiles, hence, may be interpreted in the context of a non-conservative mass transfer scenario within a binary system. In such systems, the ejection of material through the outer Lagrangian points can lead to the formation of asymmetric and shell-like circumstellar structures. The resulting distribution of shells in the CSM is expected to be highly anisotropic and dependent on the orbital phase and orientation of the companion star at the time of ejection.

Consequently, our line of sight may intersect different regions of the CSM at different epochs, resulting in apparent variations in the bulk motion of the absorbing/emitting material. In particular, epochs showing redshifted absorption components may correspond to shells moving away from the observer, potentially formed when the companion star was located on the far side of the system. Conversely, blueshifted features may trace clumps ejected in our direction, likely originating when the companion was on the near side. This interpretation supports a scenario in which the CSM is shaped by episodic, directional mass-loss events driven by binary interaction, rather than a spherically symmetric wind.

\begin{figure*}    
    \includegraphics[scale=0.3]{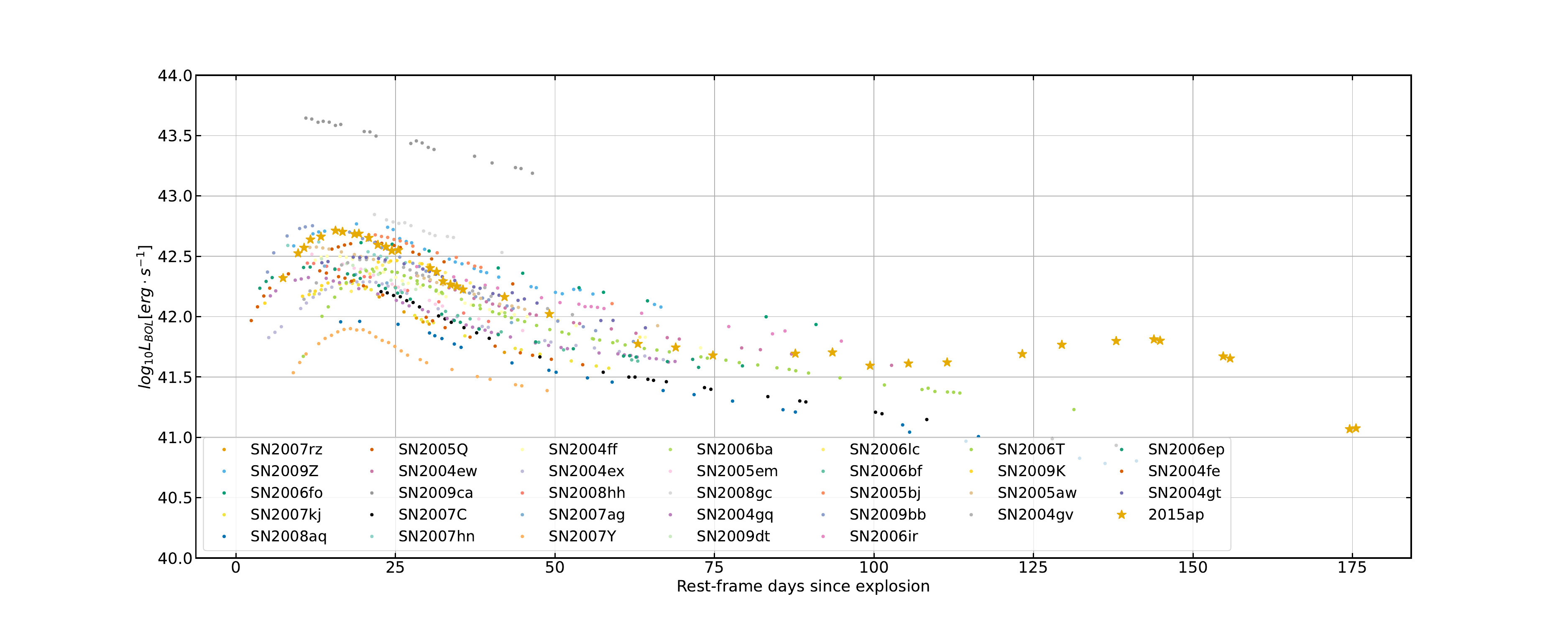}
    \caption{The figure shows the comparison of the bolometric light curve of SN 2015ap with other SESNe from the Carnegie SN survey. We report the details of the analysis in Sect.~\ref{sec:bolometric}.}\label{fig:SEbol}
\end{figure*}

\begin{figure*}    
    \includegraphics[scale=0.35]{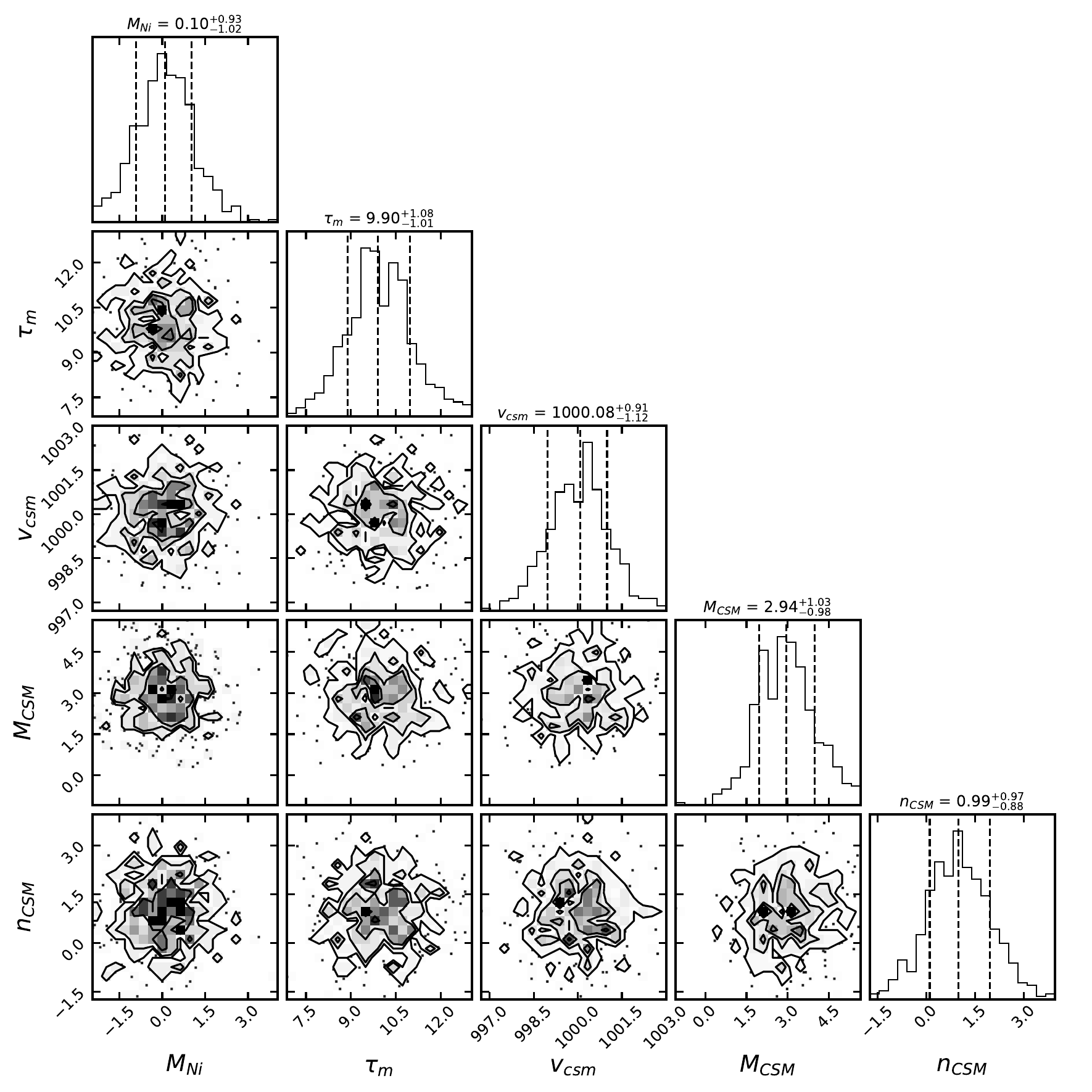}
    \caption{The figure shows the corner plot from the Bayesian fitting of the SN 2015ap's bolometric light curve and the Ni+CSM powered SN model.}\label{fig:arnettCSMcorner}
\end{figure*}
\section{Bolometric light curve}\label{sec:bolometric}
The bolometric light curve represents an estimate of the total energy output of the SN across all wavelengths and is therefore a direct probe of the intrinsic luminosity of the event. Its reconstruction is essential to derive key physical parameters such as the synthesised $^{56}Ni$, the ejected mass, and the kinetic energy of the explosion. In particular, comparing the observed bolometric light curve with that predicted by models -- whether powered by radioactive decay, central engine activity, or interaction with circumstellar material -- allows us to distinguish between different progenitor scenarios. Additionally, the shape of the bolometric curve — including the peak time, width, and decline rate — places constraints on physical timescales such as diffusion and transparency, contributing to a quantitative understanding of the evolution of the explosion.
The quasi-bolometric light curve was derived using the \texttt{superbol} code \citep{Nicholl_2018}, fitting blackbody models to extinction-corrected data. Data selection, cosmological corrections, mapping to a common time grid, blackbody fits to account for flux not covered by the observations, and light curve integration are all handled by the interactive code that is executed through a sequence of command-line prompts and displays. 
We used the photometry data from UV, UVW1, UVW2, UVM2 bands, to the near-infrared I band, which allows the integration of the quasi-bolometric luminosity over the wavelength range $1800-9000 \AA$. We measured a peak flux of $\log F = 42.72 \pm 0.02$ in agreement with the result in \citep{Prentice_2018}. 
We noted that the bolometric luminosity (first panel in Fig. \ref{fig:superbol}) shows a prominent bump in the later evolution of the light curve, which is not present in \citet{Aryan2021} or in \citep{Prentice_2018}. However, the presence of the bump is in agreement with the colour evolution, already presented in sec. \ref{sec:spectra}, suggesting a rebrightening due to interaction, probably with another CSM shell. This is also reflected in the black-body radius evolution, which by the time of the last bump, appears to expand up to $\approx 10^{17}$ cm.

To place SN 2015ap in a broader context, we compare the bolometric light curve to those of other well-studied type Ic-BL SNe \citep{2018Taddia, 2018A&A...609A.135S}. As shown in Figure \ref{fig:SEbol}, the trend of SN 2015ap is intermediate between SN 2002ap and SN 1998bw, suggesting a moderately energetic explosion. The ejecta mass and kinetic energy are also within the typical range for Ic-BL SNe not associated with gamma-ray bursts, indicating that while SN 2015ap is luminous, it does not necessarily require an extreme progenitor. This comparison supports the hypothesis that SN 2015ap belongs to the class of engine-driven Ic-BL SNe without GRB detection.

\begin{table}
\centering
\begin{tabular}{lc}\hline
 & Ni+CSM   \\ \hline
$M_{Ni}$ & 0.10$^{+1.0}_{-0.9}$ M$_{\odot}$ \\
$\tau_m $& 9.90$^{+1.0}_{-1.1}$ days\\
$v_{csm}$ & 1000.08$^{+1.1}_{-0.9}$  km s$^{-1}$\\
$M_{CSM}$ & 2.9$^{1.0}_{1.0}$ M$_{\odot}$ \\
$n_{CSM}$ & 1.0$^{+0.9}_{-1.0} $\\ \hline
\end{tabular}
\caption{The table presents the values estimated by the Bayesian fitting of the SN 2015ap's bolometric light curve.}
    \label{tab:arnetcsmfit}
\end{table}

Bearing this in mind, we estimate $M_{Ni}$, CSM mass, $M_{CSM}$, and velocity,$v_{CSM}$, explosion energy,  $E_{exp}$, assuming a radioactive + CSM interaction powered SNe. {We modeled the bolometric light curve obtained with \texttt{superbol} using analytical prescriptions from \citet{1982Arnett} and \citet{Chatzopoulos}, which account for the radioactive 
$^{56}Ni$-decay and the interaction with circumstellar material (CSM), respectively, as the main powering sources.}
The results of the fit are presented in Table \ref{tab:arnetcsmfit}. The posterior distribution of the Bayesian fitting can be seen in Fig. \ref{fig:arnettCSMcorner}. Using best fit values we could derive an $M_{ej} = 2.36\pm0.47~M_{\odot}$ and an explosion energy $E_{exp}= (3.38\pm 0.68) \times10^{51}$ erg. 
These values are compatible within the error with the estimates of \citet{2023ApJ...955...71R}. Using a velocity at maximum light of $9000$ km/s and $k=0.07 cm^2/g$, they obtain $M_ej=2.4\pm0.9 M_{\odot} ~\text{and}~ E_k=(1.17\pm0.64)\times10^51$ erg, consistent with those reported by \citet{Aryan2021}. They also derive a $^{56}Ni$ mass of $0.11\pm0.03 M_{\odot}$ using the luminosity in the radioactive tail and in the maximum light in a Ni-powered scenario, which is in agreement with our derivation using Ni+CSM model. 

The post-peak decline of the bolometric light curve follows a rate of approximately 0.025 mag day$^{-1}$, which is broadly consistent with the decay rate of $^{56}$Co to $^{56}$Fe ($\approx 0.0098\,\text{mag}\,\text{day}^{-1}$) assuming full $\gamma$-ray trapping. The slightly faster decline suggests that the ejecta become partially transparent to gamma-rays at late times. This transition is commonly observed in stripped-envelope SNe and provides further support for the estimated ejecta mass \citep{2020Dessart, 2020Sharon}. The lack of flattening in the tail disfavors strong late-time CSM interaction or prolonged central engine activity.

\section{Discussion}
Many cases of bumps or reported modulations in the late-time evolution of SESNe can be found in the literature \citep{2016Arcavi,2015BenAmi,2019Prentice,2017Taddia,2021Jacobson-Galan}.
However, no consensus has yet been reached within the scientific community on the origin of these features. Several studies \citep[e.g.][]{2012Yoon,2015AMoriya,2016Eldridge,Milisavljevic2020,Mauerhan2018,2019Yoon,Sollerman2020,Zenati2022} attribute them to shock interaction with circumstellar material (CSM) expelled by the progenitor during its evolution. Such mass-loss episodes may be linked to binary interaction or to instabilities associated with advanced convective shell burning in massive stars. Among the proposed scenarios, Roche-lobe overflow in a binary system remains one of the most straightforward and physically plausible mechanisms to account for these late-time bumps.
Adopting an agnostic perspective, we investigate the observational features of SN 2015ap in order to explore a possible physical explanation for the light-curve evolution we observe.

\subsection{Interaction with CSM}
To verify whether the bumps were ascribable to ejecta-CSM interaction, we assume two different origins of the CSM:
\begin{itemize}
    \item multiple episodes of mass loss of a single massive star evolution during the late stages of its life;
    \item mass loss related to Roche-Lobe overflow due to binary interaction.
\end{itemize}

The ejecta--CSM interaction is one of the main mechanisms responsible for SN light-curve rebrightening \citep[e.g.,][]{Tartaglia, McDowell_2018, Suzuki_2019, 2020Sollerman, Pellegrino_2022}. However, the presence of multiple bumps with time periodicity requires an \textit{ad hoc} CSM distribution, such as concentric shells expelled by the progenitor in the final stages of its evolution.

During the undulations, SN~2015ap brightens up to $\sim 10^{42} \, \mathrm{erg\,s^{-1}}$ for approximately 8 days. Using the scaling relation $L = \frac{1}{2} M_\mathrm{CSM} v^2 / t_\mathrm{rise}$ \citep{Smith_2007, Quimby_2007, 2016Nicholl}, we estimate the CSM mass required to power a single bump to be $M_{\mathrm{CSM,bump}} \approx 3.7 \times 10^{-4} \, \mathrm{M_\odot}$, assuming an ejecta velocity $v = 11200 \, \mathrm{km\,s^{-1}}$ and a rise time $t_\mathrm{rise} = 5.4$ days. {The value of ejecta velocity is estimated in sec. \ref{modelling}, while the rising time is extrapolated by a linear fitting to early phase of the bolometric light curve.}

The average pre-explosion mass-loss rate needed to create this amount of CSM can be calculated using the relation $\dot{M}/v_w = M_\mathrm{CSM}/\Delta R$, where $v_w$ is the wind velocity and $\Delta R = v t_\mathrm{bump}$ is the radial thickness of the shell. For $v = 11800 \, \mathrm{km\,s^{-1}}$ and $t_\mathrm{bump} = 8.41$ days {(assuming it by the periodicity of the SN light curves' tail)}, this gives a mass-loss rate of approximately
\[
\dot{M} \approx 1.4 \times 10^{-3} \left( \frac{v_w}{1000\,\mathrm{km\,s^{-1}}} \right) \, \mathrm{M_\odot\,yr^{-1}}.
\]
This is significantly higher than the typical mass-loss rates observed in WR stars \citep[e.g.,][]{2020Sander}, suggesting that a steady wind alone may not be sufficient to form the CSM shells needed to explain the observed modulation in SN~2015ap.

Nested dust shells produced by colliding winds in the massive binary system WR~140 have been spectacularly revealed by \textit{JWST} imaging \citep{2022Lau}. These 17 observed shells form due to repeated dust-formation episodes every 7.93 years, triggered by the periastron passage of an O5.5fc companion star orbiting the WC7 Wolf--Rayet primary. If SN~2015ap's undulations were caused by density enhancements similar to the WR~140 shells, the binary progenitor would have needed to eject such shells roughly 400 times more frequently. A more compelling argument against this scenario comes from the ejecta velocity ($11800\,\mathrm{km\,s^{-1}}$) and the 8.4-day periodicity observed in SN~2015ap. This implies a shell separation of about 57\,AU---three orders of magnitude smaller than the $\sim 4400 \, \mathrm{AU}$ separation observed in WR~140. Therefore, the progenitor system of SN~2015ap would need to eject shells every $\sim 40$ days, which is highly unrealistic compared to the 8-year interval in WR~140.

As described by \cite{2019Yoon}, a mass loss history model that can justify this kind of evolution typically involves the presence of a companion star in a binary system, which pushes the WR star to its Eddington limit by accreting its outer envelope material. 
This can be outlined by examining the colour evolution of the SN light curve. 
The colour curve in the early stages of a SN Ib/Ic can provide information on the degree of mixing of the 
$^{56}Ni$ in the ejected materials. Specifically:
\begin{itemize}
    \item A weak mixing of the $^{56}Ni$ leads to a colour curve characterised by three phases: rapid initial reddening, a reversal toward blue due to heating of the $^{56}Ni$, and finally a return to red until the nebular phase.
    \item Strong mixing of the $^{56}Ni$, on the other hand, suppresses the reversion phase toward blue and leads to a monotonous evolution of the colour curve.
\end{itemize}

However, the CSM could be produced because a binary system in a Roche-Lobe phase, due to non-conservative mass transfer, loses mass from the most external Lagrangian points (thus, producing spirals, circumstellar shells or equatorial disks).

To assess the stability of Roche lobe overflow (RLOF) in the progenitor system of SN~2015ap, we consider a binary composed of two massive stars. The donor star is assumed to have a mass of \( M_1 = 15\,M_\odot \), and the companion (accretor) a mass of \( M_2 \geq 13.8\,M_\odot \), consistent with a main-sequence or evolved OB-type star, according to the third Kepler's law. The orbital period is taken to be \( P_{\mathrm{orb}} = 8.4 \) days affected by the , matching the characteristic variability timescale observed in SN~2015ap,  assuming the orbital period is still not influenced by the change in mass of the primary star in the transition from the progenitor to the remnant compact object.

The criterion for dynamical stability of RLOF can be assessed by comparing the donor's mass-radius response, described by the parameter \( \zeta_* \), to the Roche lobe response \( \zeta_{\mathrm{RL}} \). For conservative mass transfer, the Roche lobe exponent is given approximately by \citep{Soberman1997}:

\begin{equation}
\zeta_{\mathrm{RL}} \approx 2.13 \left( 1 + \frac{M_1}{M_2} \right) = 4.44.
\end{equation}

For evolved massive stars, such as yellow or red supergiants or hydrogen-stripped Wolf-Rayet stars, the mass-radius exponent is typically \( \zeta_* \lesssim 0 \), as the stellar radius increases or stays constant during mass loss \citep{1998Vanbeveren, Ivanova2013}. Since \( \zeta_* < \zeta_{\mathrm{RL}} \), the mass transfer is expected to be dynamically unstable.

This instability may lead to rapid mass loss or the onset of a common envelope phase, depending on the evolutionary stage of the stars. The relatively short orbital period (\( \sim8.4 \) days) implies a close binary separation (\( a \sim 70\,R_\odot \)), and the donor would need to reach a radius of \( R \sim 30\,R_\odot \) to fill its Roche lobe. Such expansion is plausible in the late stages of a massive star's evolution and may result in enhanced and structured CSM.

The formation of dense CSM shells via episodic mass loss prior to the explosion may explain the light curve modulations observed in SN~2015ap. This scenario is consistent with similar ejecta-CSM interaction seen in type Ibn/IIn SNe \citep{Smith2014, Margutti2014}.

\begin{figure*}
    \centering
    \includegraphics[width=1\linewidth]{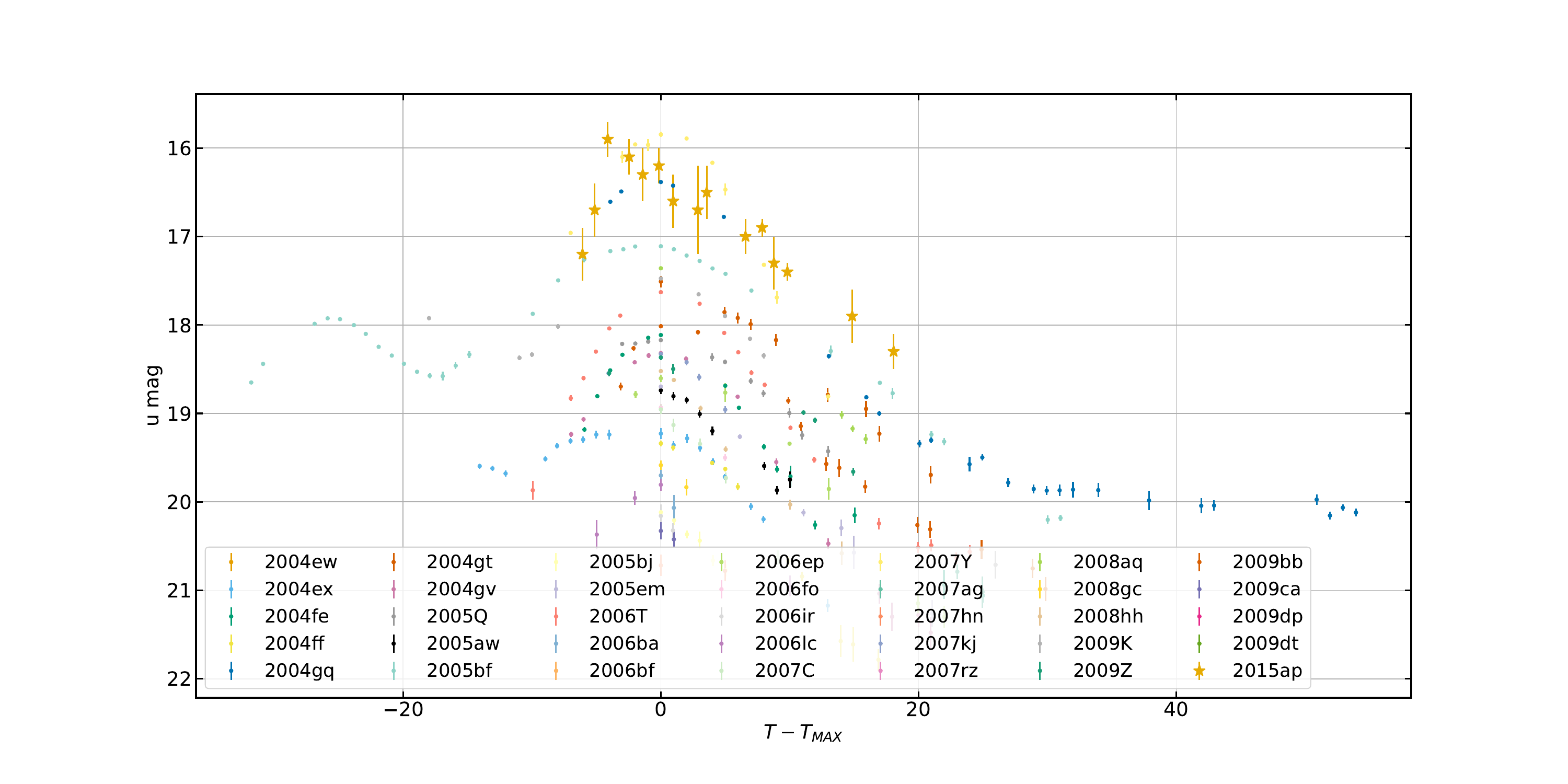}\quad \quad
    \includegraphics[width=1\linewidth]{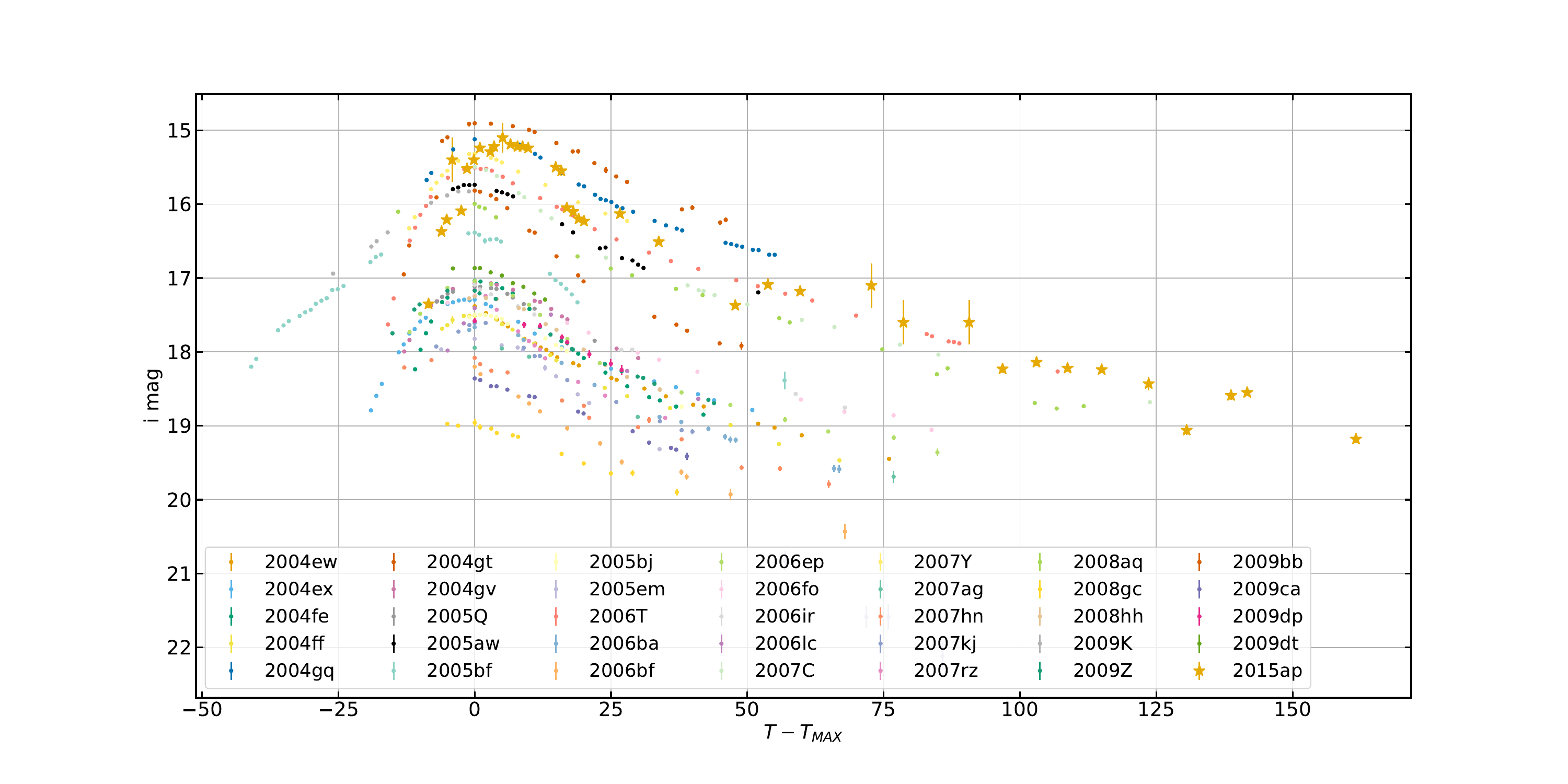}
    \caption{The figure shows the comparison of the u-band (top panel) and i-band (bottom panel) light curve of SN 2015ap with respect to other SESNe from the Carnegie SN survey. Both trends show that SN 2015ap evolves similarly to the most luminous SESNe in the sample, pointing to the presence of an additional energy source. }
    \label{fig:SE_compare}
\end{figure*}
\begin{figure}
    \centering
    \includegraphics[width=0.8\linewidth]{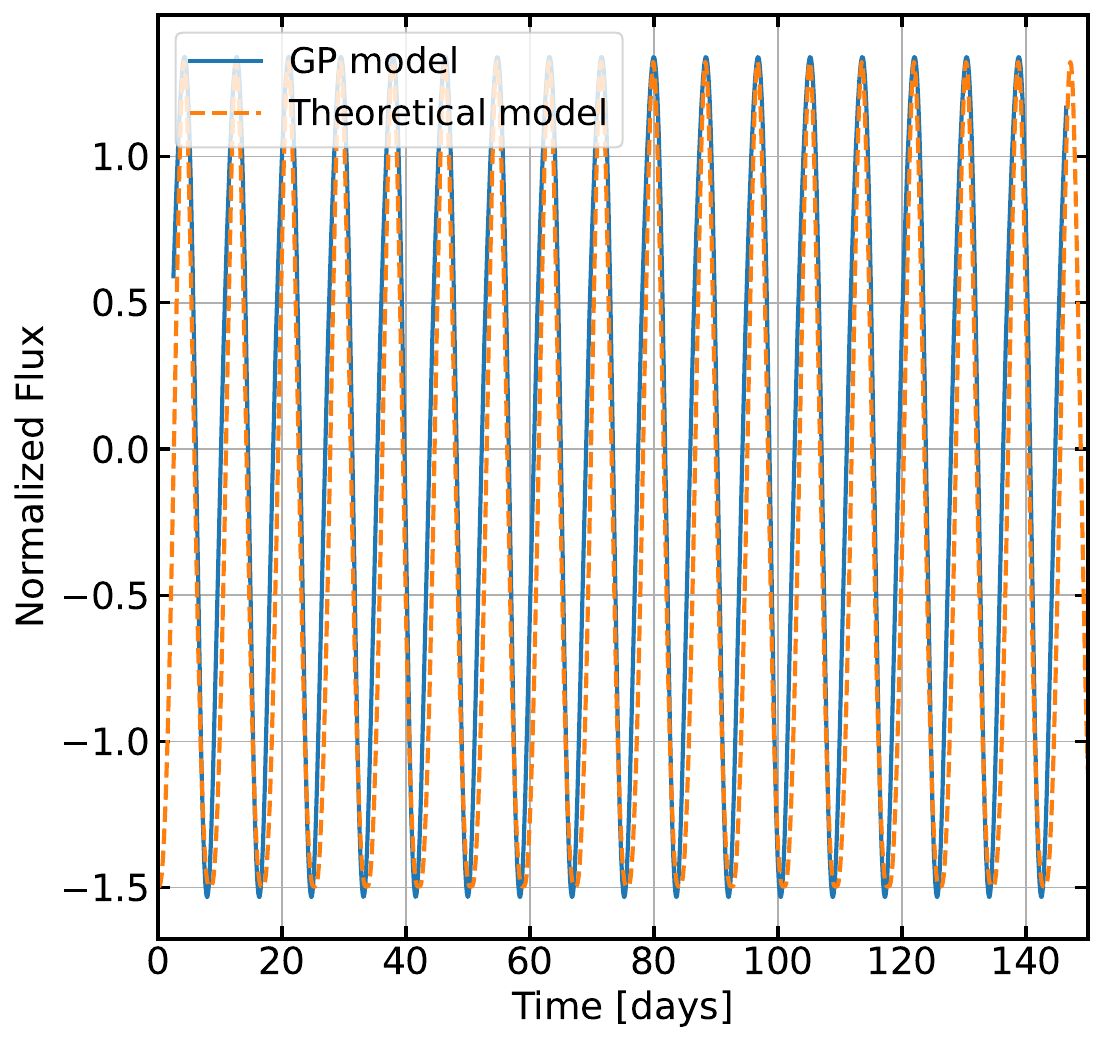}\quad \quad
    \caption{Toy model of accretion in a neutron star binary wind system, based on Bondi-Hoyle-Lyttleton formalism and modulated by a Keplerian elliptical orbit. The model flux (orange dashed line) is fitted to the normalized periodic flux extracted from the Gaussian Process (GP) analysis of SN 2015ap (blue solid line). The good agreement suggests compatibility of the observed periodicity with an accretion-driven binary wind scenario, although it is not definitive proof of proto-NS accretion.}
    \label{fig:model_theory}
\end{figure}

\subsection{Highly magnetised compact object in binary system}

Another hypothesis we propose to describe the observations is the effect on the SN observed radiation of an extra-source identified in a proto-neutron star that interacts with the survived companion of the binary system. 

The presence of a magnetar or, more generally, of a magnetised compact object scenario comes from a comparison of SN~2015ap to SESNe observed in the \textit{Carnegie Supernova Project} \citep{2018Stritzinger}. In this context, SN~2015ap stands out as one of the most luminous objects in both the $u$- and $i$-bands, see Figure \ref{fig:SE_compare}, positioning itself at the high-luminosity end of the SESN distribution. This level of brightness, particularly in bands sensitive to the blue and red ends of the optical spectrum, reinforces the idea of a central engine boosting the overall energy budget. The unusually strong $u$-band emission may directly trace high-energy input that increases the ejecta temperature or maintains heating over extended timescales, consistent with magnetar-powered emission.
The periodic rebrightenings observed in the light curve of SN~2015ap strongly suggest a binary origin. A single compact remnant cannot easily produce such regular flux modulations; instead, the recurrence implies orbital motion, pointing to the presence of a surviving companion. Using this description the origin of the blue continuum spectra observed during the rising phase and the brighter luminosity with respect to other similar transient (see Figure \ref{fig:SE_compare}) becomes clearer: it is the natural consequence of a compact object—most likely a magnetar—accreting material from its binary partner.
As shown by \citet{kasen2009, 2016Kasen}, energy injection from a magnetar can significantly boost the brightness of a SN, with deposition occurring on timescales comparable to the photon diffusion time through the ejecta. This results in a brighter light curve and a higher photospheric temperature, producing a bluer spectrum than expected from purely radioactively powered SNe.

In this scenario, we have to expect the presence of a highly magnetised NS formed at birth, which could power a SN to peak luminosities described by:

\begin{displaymath}
L_{\mathrm{peak}} \approx f\frac{E_p t_p}{t_d^2}\left[\ln\left(1+\frac{t_d}{t_p}\right)-\frac{t_d}{t_p + t_d}\right],
\end{displaymath}

where $E_p \approx 2 \times 10^{50} \left(\frac{I_{\mathrm{NS}}}{10^{45} \, \mathrm{g \, cm^2}}\right)\left(\frac{P}{10 \, \mathrm{ms}}\right)^{-2}$~erg is the rotational energy of the NS, $f$ is a correction factor, $t_d = \left(\frac{3\kappa M_{\mathrm{ej}}}{4\pi v_{\mathrm{ej}}c}\right)^{1/2}$ is the photon diffusion timescale, and $t_p = 0.44 \left(\frac{P}{10 \, \mathrm{ms}}\right)^2\left(\frac{B}{10^{14} \, \mathrm{G}}\right)^{-2}\left(\frac{I_{\mathrm{NS}}}{10^{45} \, \mathrm{g \, cm^2}}\right)\left(\frac{R_{\mathrm{NS}}}{12 \, \mathrm{km}}\right)^{-4}$~yr is the spin-down timescale. Here, $P$ denotes the spin period and $B$ is the magnetic field.

At later times, the light curve follows the spin-down luminosity:

\begin{displaymath}
L_p(t) = \frac{E_p}{t_p}\left(1 + \frac{t}{t_p}\right)^{-2}.
\end{displaymath}

By fitting the late-time light curve of SN~2015ap to $L_p(t)$, we find that a magnetised NS with $P = 2.0 \pm 1.3$~ms and $B = (6.73 \pm 0.01) \times 10^{14}$~G can reproduce the observed luminosity and decline rate. These values are consistent with those obtained from our light curve modelling using \texttt{MOSFiT} (see \autoref{modelling}). However, this scenario requires a supplementary source which is less powerful, but with a faster spin period than that considered in \citet{2020Gangopadhyay}, who modeled the bolometric light curve using a hybrid $^{56}$Ni~+~magnetar scenario, finding a $^{56}Ni$ of $\sim$0.01~M$_\odot$, ejecta mass of $\sim$3.75~M$_\odot$, and magnetar spin period of 25.8~ms with a magnetic field strength of $2.8 \times 10^{15}$~G. 
These differences may arise from several factors, including the time range over which the models are constrained and the assumptions regarding energy sources. Our model focuses on the late-time light curve, where magnetar spin-down is expected to dominate, whereas \citet{2020Gangopadhyay} fit the full bolometric evolution, including the early-time peak, where contributions from shock breakout, diffusion, or CSM interaction may play a role. The higher inferred magnetic field in their model likely reflects the need to power a brighter early light curve over a shorter timescale, while our fit suggests a longer-lived, less extreme magnetar contributing at later phases.


Rapidly rotating NSs in binaries can also accrete material from their companions, a scenario which aligns with the behaviour of X-ray binaries, where episodic accretion onto a magnetised neutron star produces bursts of high-energy radiation. In the context of SN~2015ap, these bursts could be emitted as X-rays and reprocessed by the inner layers of the SN ejecta into optical and ultraviolet light. Such reprocessed radiation would naturally enhance the luminosity in the blue and UV bands, contributing to the observed early-time colour excess.

We used a toy model for modelling the accretion process in a X-ray binary-like event (see Figure \ref{fig:model_theory}). This assumes a physical simplification of the accretion process on a neutron star (NS) in a binary wind system (not by Roche filling), and is mainly based on the Bondi-Hoyle-Lyttleton (BHL) accretion formalism, modulated by a Keplerian elliptical orbit \citep{Bondi1944, 2015Walter, 2017Mellah}. Assuming that the GP model perfectly describes the observed periodicity of SN 2015ap, we fitted the toy model for the accretion in a proto-NS/OB star binary system. Even though this cannot be considered proof that we are actually observing accretion on a proto-NS, the matching between the normalized flux from the GP and the toy model suggests that the observations are compatible with a binary wind scenario.

In SN~2015ap, the consistent profile and timescale of the bumps, along with the enhanced UV emission, are both naturally explained if a magnetar in a tight orbit accretes matter during periastron passages or through magnetic gating. The effect would be most prominent during the early phases, when the optical depth is high enough to efficiently reprocess X-rays into optical light, especially in the $u$-band. However, the higher opacity of the ejecta during earlier phases makes them visible only during later evolution of the light curves. 

\section{Conclusion}

In this study, we have presented a detailed analysis of SN~2015ap, a stripped-envelope SN exhibiting characteristics consistent with a type Ib/c event. Through photometric and spectroscopic modeling, we constrained key explosion parameters, deriving an ejecta mass of $\sim 2.2~M_\odot$, an explosion energy of $\sim 3.4 \times 10^{51}~\mathrm{erg}$, and a synthesised $^{56}$Ni mass of $\sim 0.11~M_\odot$. The observed post-peak decline rate and spectral evolution support a scenario where the light curve is primarily powered by radioactive decay with limited contribution from late-time CSM interaction. While a magnetar model can reproduce the observed features, our combined Ni+CSM modelling yields similarly good fits, highlighting the degeneracy in power source interpretations. 

Ultimately, SN~2015ap adds to the growing sample of transitional objects that bridge the gap between standard stripped-envelope SNe and more energetic explosions, underscoring the importance of early-time spectral coverage and dense photometric sampling in disentangling their nature.

Among the various models tested, the radioactive decay model powered by $^{56}$Ni+CSM interaction provides a satisfactory fit to the observed light curve. The magnetar-driven model yields a slightly better fit in the early-time rise but underestimates the late-time evolution. Based on Bayesian Information Criterion (BIC) comparisons, the $^{56}$Ni+CMS model is preferred. Thus, we are inclined to conclude that radioactive decay is the dominant power source, though we cannot entirely exclude a contribution from a weak central engine.
A very interesting outcome of our analysis is the discovery of a periodicity in the spectro-photometric evolution of SN 2015ap. This periodicity is shown mainly in the evolution of the Ha line. Therefore, we are led to interpret this behaviour as due to a particular configuration of the CSM produced by anisotropic and periodic mass loss phenomena, typical of the interaction of two stars in a binary system. Hence, we infer that the radiation from the CMS-ejecta interaction shows a periodicity of $P\approx8.4$ d. 

From the colour evolution, we also derived evidence of additional energy injection: the chromatic and thermal behaviour suggests a prolonged breakout or delayed ignition of energy, which can be explained by a central mechanism modulated by the presence of a companion  (e.g., a magnetar orbiting around the donor). However, the lack of X-ray or radio observations prevents a definitive identification of a compact object as the energy source.

Overall, SN~2015ap appears to be a transitional object, bridging the population of standard stripped-envelope SNe and more energetic engine-powered events. Its photometric modulations, CSM structure, and low ejecta mass favour a progenitor system shaped by binary evolution, highlighting the key role that binarity plays in the final stages of massive star evolution.

\section*{Data availability}

The photometric and spectroscopic data underlying this article are publicly available from previous studies \citep{Prentice_2018, 2020Gangopadhyay, Aryan2021} and via the SOUSA archive at \url{https://archive.stsci.edu/prepds/sousa/}. Additional modelling outputs generated during the current study will be shared on reasonable request to the corresponding author.

\section*{Acknowlegement}
This work was initially thought out and developed at The Unconventional Thinking Tank Conference 2022, which is supported by INAF. We thank F. Patat, S. Ascenzi and A. Papitto for fruitful discussions that improved the manuscript and that helped in the interpretation and analysis of the event. 
This work was supported by the INAF MiniGrant 2023 programme under the project title “KeNSHIRO: KNe Serendipitous Hunt In the Rubin Observatory era”.
This work was supported by the Preparing for Astrophysics with LSST Program, funded by the Heising Simons Foundation through grant 2021-2975, and administered by Las Cumbres Observatory. We acknowledges support from the National Science Foundation with grant numbers PHY-2010970 and OAC-2117997. Fabio Ragosta thanks the LSSTC Data Science Fellowship Program, which is funded by LSSTC, NSF Cybertraining Grant \#1829740, the Brinson Foundation, and the Moore Foundation; his participation in the program has benefited this work.

\bibliographystyle{mnras}
\bibliography{refs}

\bsp	
\label{lastpage}
\end{document}